\documentclass[twocolumn,
 amsmath,amssymb,
aps,
prb,
]{revtex4}

\usepackage{graphicx}
\usepackage{dcolumn}
\usepackage{bm}

\usepackage{SIunits}

\usepackage{pict2e}

\begin{document}

\title{

Influence of phonons on solid-state cavity-QED investigated using nonequilibrium Green's functions
}

\author{Gaston Hornecker} 
\author{Alexia Auff\`{e}ves}
\author{Thomas Grange}
\email{thomas.grange@neel.cnrs.fr}
 
\affiliation{Universit\'e Grenoble-Alpes, 38000 Grenoble, France}
\affiliation{CNRS, Institut N\'{e}el, "Nanophysique et semiconducteurs" group, 38000 Grenoble, France}

\date{\today}

\begin{abstract}
The influence of electron--phonon interactions on the dynamics of a quantum dot coupled to a photonic cavity mode is investigated using a nonequilibrium Green's function approach.
Within a polaron frame, the self-consistent-Born approximation is used to treat the phonon-assisted scattering processes between the quantum dot polaron and the cavity.
Two-time correlators of the quantum dot-cavity system are calculated by solving the Kadanoff-Baym equations, giving access to photon spectra and photon indistinguishability.
The non-Markovian nature of the interaction with the phonon bath is shown to be very accurately described by our method in various regime of cavity-quantum electrodynamics (cavity-QED).
The indistinguishability of the emitted photons emitted at zero temperature are found to be in very good agreement with a previously reported exact diagonalization approach [Phys.~Rev.~B~87,~081308~(2013)]. Besides, our method enables the calculations of photon indistinguishability at finite temperatures and for strong electron-phonon interactions.
More generally, our method opens new avenues in the study of open quantum system dynamics coupled to non-Markovian environments.
\end{abstract}

\maketitle


\section{Introduction}

Coupling a solid-state artificial atom to an optical cavity has recently attracted considerable interests, motivated by possible applications of cavity quantum electrodynamics (cavity-QED) in photonics, metrology and quantum information
\cite{o2009photonic,hennessy2007quantum,englund2007controlling,kasprzak2010up,ota2015vacuum}.
Quantum dot (QD)--cavity systems have enabled great progress in the development of on-demand sources of indistinguishable single photons
 \cite{santori2002indistinguishable,patel2010two,lettow2010quantum,gazzano2013bright,he2013demand,muller2014demand,
 monniello2014indistinguishable,sipahigil2014indistinguishable,wei2014deterministic,PhysRevB.92.161302,somaschi2015near,loredo2016scalable}. 
Beyond 
the picture of an ideal cavity-QED system,
couplings to phonons have been shown to play a crucial role theoretically \cite{wilson2002quantum,kaer2010non,hohenester2010cavity,roy2011phonon,hughes2011influence,roy2015spontaneous} and experimentally \cite{hohenester2009phonon,ota2009impact,majumdar2011phonon,calic2011phonon,madsen2013measuring,valente2014frequency,Portalupi2015Bright,hopfmann2015compensation,muller2015ultrafast}.
A non-Markovian model for the dissipation in the phonon reservoir is mandatory to provide a general description of the various phenomena involving phonons in these systems. Indeed the phonon reservoir stores information on a time scale corresponding to the
sound propagation time
of the generated acoustic waves out of the QD.
These non-Markovian effects appear to be particularly crucial when studying the temporal coherences involved in the emission spectrum and in the photon indistinguishability.

Various methods have been used to investigate the memory effects induced by phonons in the quantum dynamics of quantum dot-cavity systems.
Second-order perturbation theory within the time convolutionless (TCL) approach has been used to investigate non-Markovian effects for various regimes \cite{kaer2010non, roy2011phonon, kaer2012microscopic,kaer2014decoherence}.
Yet, an exact diagonalization approach reported at 0K \cite{kaer2013microscopic} 
has evidenced the failure of such finite-order TCL approaches in predicting the indistinguishability of the emitted photons in the QD-cavity strong coupling regime or in the large cavity linewidth limit. However, such an exact diagonalization method, which involves a truncation of the full Hilbert space, is computationally tractable only at zero temperature and for weak electron-phonon interaction. A more tractable approach to account accurately for the non-Markovian influence of phonons in the various regimes of solid-state cavity-quantum electrodynamics (cavity-QED) is still lacking.

Here we present a nonequilibrium Green's function (NEGF) approach to investigate the dynamics of a QD--cavity system interacting with phonons.
Within a polaron frame, the phonon-electron-photon scattering terms are treated within the self-consistent Born approximation. This allows to account for an infinite number of phonon-assisted scattering processes, beyond existing finite-order perturbation approaches. The two-time Green's functions are computed by solving the Kadanoff-Baym equations. We report calculations of the non-Markovian dynamics of a QD-cavity system interacting with a phonon bath, as well as photon emission spectrum and indistinguishability of the emitted photons. The results are shown to be in very good agreement with exactly solvable limits as well as an exact diagonalization approach at 0K \cite{kaer2013microscopic} . Our approach is shown to provide accurate and efficient simulations of the dynamics of quantum dot-cavity systems interacting with phonons.

\section{Theory}
\subsection{System Hamiltonian}
\begin{figure}
\begin{centering}
\includegraphics[width=0.45\textwidth]{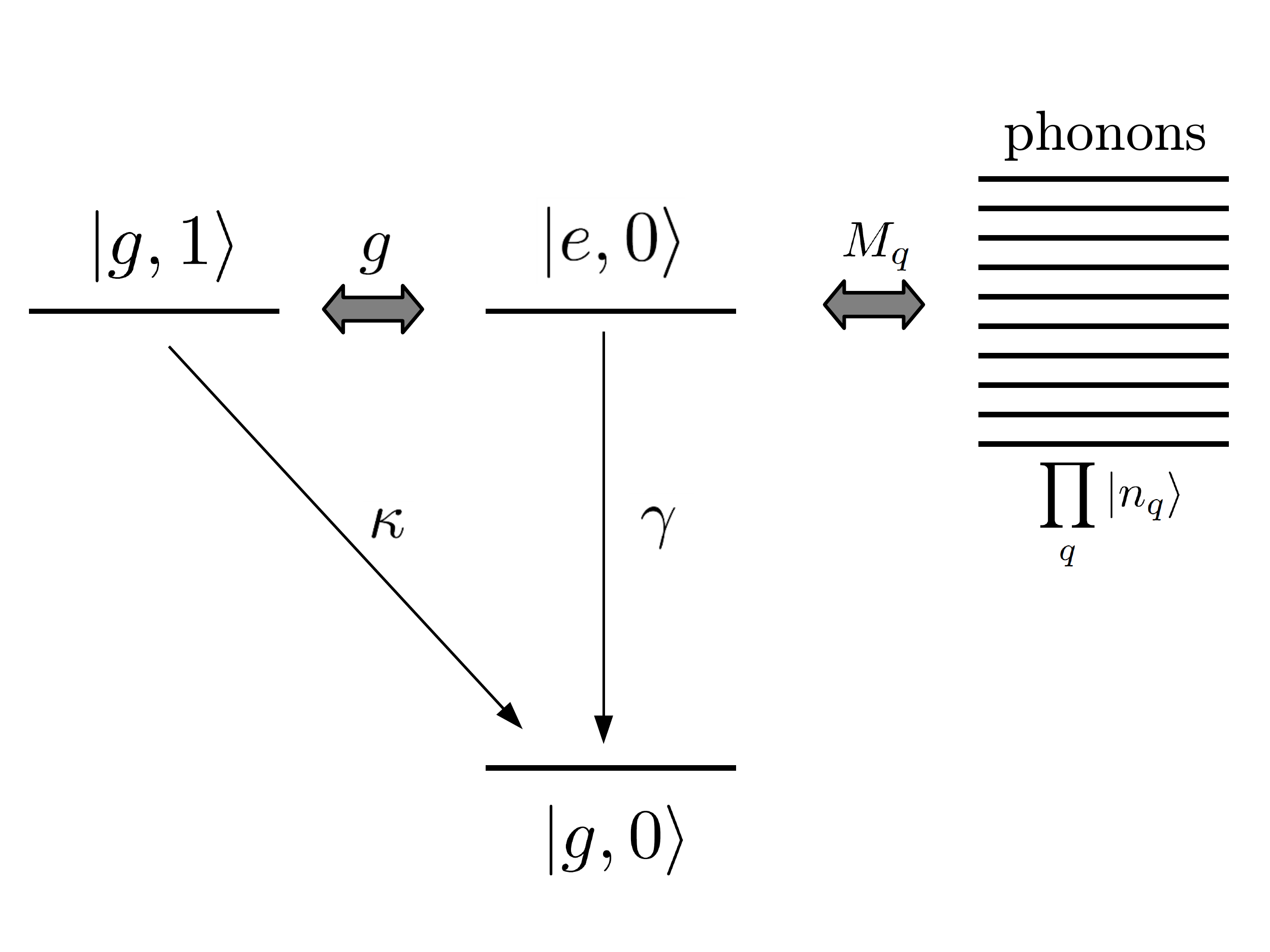}
\end{centering}
\caption{Level diagram for the solid-state cavity-QED system interacting with phonons. The 3 lowest energy levels of the Jaynes-Cummings ladder are represented with the notation $|g/e, N\rangle$, in which $|g/e\rangle$ is the two-level system ground/excited state and $|N\rangle$ is the cavity Fock state. The electronic states are coupled to a phonon continuum, whose states $\prod_q|n_q\rangle$ are tensorial products of the $|n_q\rangle$ phonon Fock states of wavevector $q$.}
\label{schema_levels}
\end{figure}

We consider an electronic two-level system (TLS) -- such as a quantum dot (QD) -- interacting simultaneously with (i) a quantized mode of an optical cavity and (ii) a phonon continuum (Fig.~\ref{schema_levels}). Within the rotating-wave approximation, the Hamiltonian of the modeled system reads:
\begin{equation}
\begin{split}
\hat{H} = &\hbar \omega_{\text{e}} \hat{c}^{\dagger}\hat{c} + \hbar \omega_{\text{cav}} \hat{a}^{\dagger}\hat{a} + \hbar g(\hat{c}\hat{a}^{\dagger}+\hat{c}^{\dagger}\hat{a}) \\
& + \hat{c}^{\dagger}\hat{c} \sum_q M_q (\hat{b}^{\dagger}_q+\hat{b}_q) + \sum_q \hbar\omega_q \hat{b}^{\dagger}_q\hat{b}_q
\end{split}
\label{hamiltonian}
\end{equation}
where $\hat{c}^{\dagger}$ and $\hat{c}$ are the fermionic creation and annihilation operators of the electronic TLS (ground and excited states $|g\rangle$ and $|e\rangle$ respectively), with frequency $\omega_{\text{e}}$; $\hat{a}^{\dagger}$ and $\hat{a}$ are the photon ladder operators of the cavity, with frequency $\omega_{\text{cav}}$; $g$ is the electron-cavity coupling strength;
$\hat{b}^{\dagger}_q$ and $\hat{b}_q$ are the phonon ladder operators of wavevector $q$; $M_q$ is the electron-phonon coupling strength.  The cavity photon Fock states are noted $|N\rangle$, while the phonon Fock states of wavevector $q$ are noted $|n_q\rangle$.

In the limit of vanishing exciton-photon couling strength $g=0$, the above Hamiltonian reduces to the independent boson model, which can be solved by a polaron transformation \cite{mahan2000many}.
In the following, such polaron transformation is used. The physical motivation is to account exactly for the cavity-QED effects that involves the zero-phonon line, and to treat perturbatively the coupling between the phonon sidebands and the cavity.

\subsection{Polaron transformation}

We start by applying the unitary transformation that diagonalizes the independent boson Hamiltonian \cite{mahan2000many}.
The polaron transformation reads for an operator $\hat{O}$:

\begin{equation}
\hat{O}' = e^{\hat{S}} \hat{O} e^{-\hat{S}}
\end{equation}
with 
\begin{equation}
\hat{S} = \hat{c}^{\dagger}\hat{c} \sum_q \frac{M_q}{\hbar\omega_q} (\hat{b}^{\dagger}_q-\hat{b}_q)
\end{equation}
This transformation shifts the position of the lattice when the QD is excited in its new equilibrium position, i.e. in a polaron state. The creation and annihilation operators are transformed as $\hat{c}^{(\dagger)} \rightarrow \mathcal{D}^{(\dagger)} \hat{c}^{(\dagger)}$.
The transformed Hamiltonian reads:
\begin{equation}
\begin{split}
\hat{H}'= & (\hbar \omega_{\text{e}}-\Delta_p) \hat{c}^{\dagger}\hat{c} + \hbar \omega_{\text{cav}} \hat{a}^{\dagger}\hat{a} + \sum_q \hbar\omega_q \hat{b}^{\dagger}_q\hat{b}_q \\
& + \hbar g \left[ \hat{c}\hat{a}^{\dagger}\mathcal{D}+\hat{c}^{\dagger}\hat{a}\mathcal{D}^{\dagger} \right] ,
\end{split}
\end{equation}
where $\hat{\mathcal{D}}$ is the phonon displacement operator corresponding to the polaron formation
\begin{equation}
\hat{\mathcal{D}} =  \exp \left[-\sum_q \frac{M_q}{\hbar\omega_q} (b^{\dagger}_q-b_q) \right]
\end{equation}
and $\Delta_p$ the polaron energy shift:
\begin{equation}
\Delta_p = \sum_q \frac{\vert M_q \vert ^2}{\hbar\omega_q}.
\end{equation}
A renormalized exciton energy is defined due to this polaron shift as 
\begin{equation}
\tilde{\omega}_{\text{e}} = \omega_{\text{e}}  - \Delta_p .
\end{equation}
We then separate the transformed Hamiltonian into three terms:
\begin{equation}
\hat{H}' = \hat{H}_0  +  \hat{H}_{\text{ph}} + \hat{H}_{\text{int}} ,
\end{equation}
where the first term reads:
\begin{equation}
\hat{H} _0= (\hbar \omega_{\text{e}}-\Delta_p) \hat{c}^{\dagger}\hat{c} + \hbar \omega_{\text{cav}} \hat{a}^{\dagger}\hat{a} +  \hbar g \langle \hat{\mathcal{D}} \rangle (\hat{c}\hat{a}^{\dagger}+\hat{c}^{\dagger}\hat{a} ) .
\end{equation}
It describes the coupling between the QD polaron and the cavity without the exchange of phonons, from which originates the zero-phonon line. This coupling is renormalized by a factor $\langle \hat{\mathcal{D}} \rangle$, which is the expectation value of the phonon displacement in a thermal state \cite{mahan2000many}:
\begin{equation}
\langle \hat{\mathcal{D}} \rangle = e^{-\varphi(0)/2},
\end{equation}
\begin{equation}
\varphi(\tau) = \sum_q \frac{M_q^2}{\hbar^2\omega_q^2} \left[ N_{\omega_q}e^{i\omega_q \tau} + (N_{\omega_q}+1)e^{-i\omega_q \tau} \right],
\end{equation}
where $N_{\omega}$ is the Bose-Einstein factor at frequency $\omega$.
The second term corresponds to non-interacting phonons:
\begin{equation}
\hat{H}_{\text{ph}}=\sum_q \hbar\omega_q \hat{b}^{\dagger}_q\hat{b}_q
\end{equation}
The third term describes simultaneous photon-electron-phonon scattering (i.e. phonon-assisted exchange between the QD polaron and the cavity):
\begin{equation}
\hat{H}_{\text{int}} = \hbar g (\hat{c}\hat{a}^{\dagger} \hat{\delta \mathcal{D}}+\hat{c}^{\dagger}\hat{a} \hat{\delta \mathcal{D}}^{\dagger} ), 
\label{Hint}
\end{equation}
where
\begin{equation}
\hat{\delta \mathcal{D}} = \hat{\mathcal{D}}-\langle \hat{\mathcal{D}} \rangle
\end{equation}
is the difference between the displacement operator and its thermal expectation value.

Such polaron transformation was already used previously to treat QD--cavity systems in presence of phonon couplings \cite{wilson2002quantum,kaer2012microscopic}. Second-order perturbation theories \cite{wilson2002quantum,kaer2012microscopic} have been applied to treat this term.
In the following we go beyond these existing approaches and treat this interacting term within the self-consistent Born approximation. To this purpose, we make use of the nonequilibrium Green's function (NEGF) formalism.


\subsection{Nonequilibrium Green's functions}

To calculate the non-Markovian dynamics of the QD--cavity system in presence of the phonon environment, we use the nonequilibrium Green's function (NEGF) formalism \cite{haug2008quantum}.
Nonequilibrium Green's function are usually defined for single particles within many-body systems, and allows for the perturbative treatment of many-body interactions.
Here in contrast we consider Green's functions (GFs) within the full Hilbert space of a small-dimension system -- the coupled quantum-dot--cavity system.
Perturbation theory is then used to treat the interactions with an external environment -- i.e. the phonon bath.
As discussed in annex, this approach requires slightly different definitions of the GFs with respect to the usual NEGF formalism \cite{haug2008quantum}.

We consider the basis formed by the states $\left\{ |i\rangle \right\} = \left\{ |g\rangle, | e \rangle \right\} \otimes \{ |N\rangle \} $, i.e. the tensor product of the electronic two-level system and the cavity Fock states. 
$\hat{\Psi}^{(\dagger)}_i$ represents the fermionic annihilation (creation) operator in the state $|i\rangle $.
We define the lesser and retarded GFs respectively by
\begin{subequations}
\begin{equation}
G_{i,j}^<(t_1,t_2) = i \langle \hat{\Psi}^{\dagger}_j (t_2) \hat{\Psi}_i(t_1) \rangle
\end{equation}
\begin{equation}
G_{i,j}^{R}(t_1,t_2) = -i\Theta(t_1-t_2)\langle \hat{\Psi}_i(t_1) \hat{\Psi}^{\dagger}_j (t_2)  \rangle_{\text{vac}}
\end{equation}
\label{DefGF}
\end{subequations}
in which the retarded GF involves expectation values $\langle \rangle_{\text{vac}}$ on the electron-photon vacuum state. 
As shown in appendix, the GFs defined above fulfill the same equations of motion as standard GFs, namely the Kadanoff-Baym equations:

\begin{widetext}

\begin{subequations}
\begin{equation}
i \hbar \frac{\partial}{\partial t_1} \hat{G}^R(t_1,t_2) - \hat{H_0}(t_1)  \hat{G}^R(t_1,t_2) =  \hbar\delta(t_1-t_2)\hat{\mathrm{I}} + \int dt \hat{\Sigma}^R(t_1,t)\hat{G}^R(t,t_2)
\end{equation}

\begin{equation}
-i  \hbar \frac{\partial}{\partial t_2} \hat{G}^R(t_1,t_2) -  \hat{G}^R(t_1,t_2)\hat{H_0}(t_2) =  \hbar\delta(t_1-t_2)\hat{\mathrm{I}} + \int dt \hat{G}^R(t_1,t)\hat{\Sigma}^R(t,t_2)
\end{equation}

\begin{equation}
i  \hbar \frac{\partial}{\partial t_1} \hat{G}^<(t_1,t_2) - \hat{H_0}(t_1)  \hat{G}^<(t_1,t_2) = \int dt \left[ \hat{\Sigma}^R(t_1,t)\hat{G}^<(t,t_2) + \hat{\Sigma}^<(t_1,t)\hat{G}^A(t,t_2) \right]
\end{equation}

\begin{equation}
-i  \hbar \frac{\partial}{\partial t_2} \hat{G}^<(t_1,t_2) - \hat{G}^<(t_1,t_2)  \hat{H_0}(t_2)  = \int dt \left[ \hat{G}^R(t_1,t)\hat{\Sigma}^<(t,t_2) + \hat{G}^<(t_1,t)\hat{\Sigma}^A(t,t_2) \right]
\end{equation}
\end{subequations}
\end{widetext}
where $\hat{\Sigma}^<$, $\hat{\Sigma}^R$ and  $\hat{\Sigma}^A$ are respectively the lesser, retarded and advanced self-energies. In the following, the nonequilibrium GFs are calculated within the polaron frame, and the above self-energies account for the $\hat{H}_{\text{int}}$ interacting term within the polaron picture. Their calculation is discussed below.

\subsection{Self-energies}
\subsubsection{Non-Markovian self-energies due to phonons}

In the following the expectation values from the phonon bath are assumed to be  given by their thermal expectations.
Under this assumption, the expectation value of the interacting term vanishes, i.e. $\langle H_{\text{int}} \rangle_{\text{bath}} =0$.
The lowest-order non-vanishing terms are of second order with respect to $H_{\text{int}}$ and corresponds to the first-order Born approximation which describes a single scattering process. To describe up to an infinite number of scattering processes, we go beyond and use the self-consistent Born approximation (SCBA). The lesser self-energy corresponding to the interacting term in the polaron frame reads
\begin{equation}
\hat{\Sigma}(t_1,t_2) = \langle \hat{H}_{\text{int}}(t_2) \hat{G}^<(t_1,t_2) \hat{H}_{\text{int}}(t_1)  \rangle_{\text{bath}}.
\end{equation}
Plugging in the expression of $\hat{H}_{\text{int}}$ (Eq.~\ref{Hint}) leads to
\begin{equation}
\begin{split}
& \hat{\Sigma}^<(t_1,t_2) = \hbar^2 |g|^2 \big[\\
&  F_{\text{m}}(t_2-t_1) \hat{c}\hat{a}^{\dagger}\hat{G}^<(t_1,t_2) \hat{c}\hat{a}^{\dagger} \\
& + F_{\text{p}}(\tau)(t_2-t_1) \hat{c}\hat{a}^{\dagger}\hat{G}^<(t_1,t_2) \hat{c}^{\dagger}\hat{a} \\
& + F_{\text{p}}(\tau)(t_2-t_1) \hat{c}^{\dagger}\hat{a}\hat{G}^<(t_1,t_2) \hat{c}\hat{a}^{\dagger} \\
& +  F_{\text{m}}(t_2-t_1) \hat{c}^{\dagger}\hat{a}\hat{G}^<(t_1,t_2) \hat{c}^{\dagger}\hat{a} \big] ,\\ 
\end{split}
\label{sigmalesser}
\end{equation}
where $F_{\text{m/p}}$ are polaron Green's functions defined by

\begin{subequations}
\begin{equation}
 F_{\text{p}}(\tau) =  \langle \delta \hat{\mathcal{D}}(t+\tau)\delta \hat{\mathcal{D}}^+(t) \rangle_{\text{bath}} = e^{-\varphi(0)}\left[ e^{\varphi(\tau)}-1 \right] ,
\end{equation}
\begin{equation}
F_{\text{m}}(\tau)  =  \langle \delta \hat{\mathcal{D}}(t+\tau)\delta \hat{\mathcal{D}}(t) \rangle_{\text{bath}} = e^{-\varphi(0)} \left[ e^{-\varphi(\tau)}-1 \right] .
\end{equation}
\end{subequations}

Similarly, the retarded self-energy reads
\begin{equation}
\hat{\Sigma}^R(t_1,t_2) = \langle \hat{H}_{\text{int}}(t_1) \hat{G}^R(t_1,t_2) \hat{H}_{\text{int}}(t_2) \rangle_{\text{bath}} ,
\end{equation}
which leads to
\begin{equation}
\begin{split}
& \hat{\Sigma}^R(t_1,t_2) = \hbar^2 |g|^2 \big[\\
&  F_{\text{m}}(t_1-t_2) \hat{c}\hat{a}^{\dagger}\hat{G}^R(t_1,t_2) \hat{c}\hat{a}^{\dagger} \\
& + F_{\text{p}}(t_1-t_2) \hat{c}\hat{a}^{\dagger}\hat{G}^R(t_1,t_2) \hat{c}^{\dagger}\hat{a} \\
& + F_{\text{p}}(t_1-t_2) \hat{c}^{\dagger}\hat{a}\hat{G}^R(t_1,t_2) \hat{c}\hat{a}^{\dagger} \\
& +  F_{\text{m}}(t_1-t_2) \hat{c}^{\dagger}\hat{a}\hat{G}^R(t_1,t_2) \hat{c}^{\dagger}\hat{a} \big] .\\
\end{split}
\label{sigmaretarded}
\end{equation}

In the numerical implementation, the non-Markovian phonon self-energies are computed for correlation times up to a maximum value $\tau_{\text{c}}^{\text{m}}$, i.e. only self-energies $\hat{\Sigma}(t+\tau,t)$ with $\tau < \tau_{\text{c}}^{\text{m}}$ are computed. 
In the full calculations shown below, $\tau_{\text{c}}^{\text{m}}$ is taken large enough so that the computed physical quantities reach converged values.

\subsubsection{Markovian self-energies}
In addition to the above non-Markovian self-energies arising from the interaction with phonons, Markovian dissipative terms are used to describe other damping processes: the cavity damping rate $\kappa$, the QD population decay rate $\gamma$, and the QD pure dephasing rate $\gamma^{*}$.
The corresponding retarded and lesser self-energies read:

\begin{equation}
\begin{split}
\hat{\Sigma}^R(t_1,t_2) & =  -i\delta(t_1-t_2)\Theta(t_1-t_2)(\gamma+\gamma^*) \hat{c}^{\dagger}\hat{c} \\
  & -i\delta(t_1-t_2)\Theta(t_1-t_2)\kappa\hat{a}^{\dagger}\hat{a} ,
\end{split}
\end{equation}

\begin{equation}
\hat{\Sigma}^<(t_1,t_2) = 
\delta(t_1-t_2) \gamma^* \hat{c}^{\dagger}\hat{c} \hat{G}(t_1,t_2) \hat{c}^{\dagger}\hat{c} .
\end{equation}
In the absence of non-Markovian phonon self-energies, these self-energies injected in the Kadanoff-Baym equations give the same Lindblad equation for the density matrix ($\hat{\rho}(t) = -i\hat{G}^<(t,t)$) as in Refs.~\onlinecite{auffeves2010controlling} and \onlinecite{grange2015cavity}.



\subsection{Exciton-phonon coupling terms}

The theory presented above is applied below to the case of a self-assembled QD coupled to bulk acoustic phonons.
We consider the coupling between the fundamental exciton (modeled as an electron-hole pair wavefunction in their respective ground state) and the 3D bulk longitudinal acoustic (LA) phonons.
Note that coupling of acoustic phonons to higher exciton modes are not considered here, as the induced dephasing effects can be accounted within the Markovian pure dephasing term $\gamma^*$ \cite{muljarov04,grange2009decoherence}.
The coupling term between the fundamental exciton and the LA-phonons due to deformation potential reads:
\begin{equation}
M_q = \sqrt{\frac{\hbar q}{2 c_s \rho_m V}}  \left[D_e \langle \phi_e | e^{-qr_e}| \phi_e \rangle + D_h \langle \phi_h |e^{-qr_h} | \phi_h \rangle \right]
\end{equation}
where $q$ is the phonon wavevector, $c_s$ is the speed of sound, $\rho_m$ the mass density, $V$ the crystal volume, $D_e$ and $D_h$ the deformation potential for the electrons and holes, and $|\phi_{e/h} \rangle$ the electron/hole envelope wavefunctions.
Bulk GaAs parameters of $c_s = 5110$~m.s$^{-1}$ and $\rho_m = 5370$~kg.m$^{-3}$ are used in the following.
For simplicity, identical envelope wavefunctions are assumed for electrons and holes with isotropic Gaussian shape:
\begin{equation}
\phi_{e/h}(r) =\frac{e^{-r^2/2\sigma^2}}{\pi^{3/4} \sigma^{3/2}} ,
\end{equation}
where the $\sigma$ is a confinement length.
This leads to a exciton-phonon coupling term of
\begin{equation}
M_q = D \sqrt{\frac{\hbar q}{2 c_s \rho_m V}}  e^{-\sigma^2 q^2/4} ,
\end{equation}
where $D=D_e+D_h$ is the exciton deformation potential. The corresponding spectral density reads:

\begin{equation}
J(\omega) = \sum_q M_q^2 \delta(\hbar\omega - \hbar\omega_q) = \frac{\omega^3}{4\pi^2  \rho_m c_s^5  }
D^2 e^{-\sigma^2 q^2/2}.
\end{equation}
Note that such $\omega^n$ dependence with $n>2$ implies a non-zero value of $\langle \hat{\mathcal{D}} \rangle^2$ the weight  of the zero-phonon line (ZPL).

\section{Results and discussion}

\begin{figure*}
\begin{centering}
\includegraphics[width=0.45\textwidth]{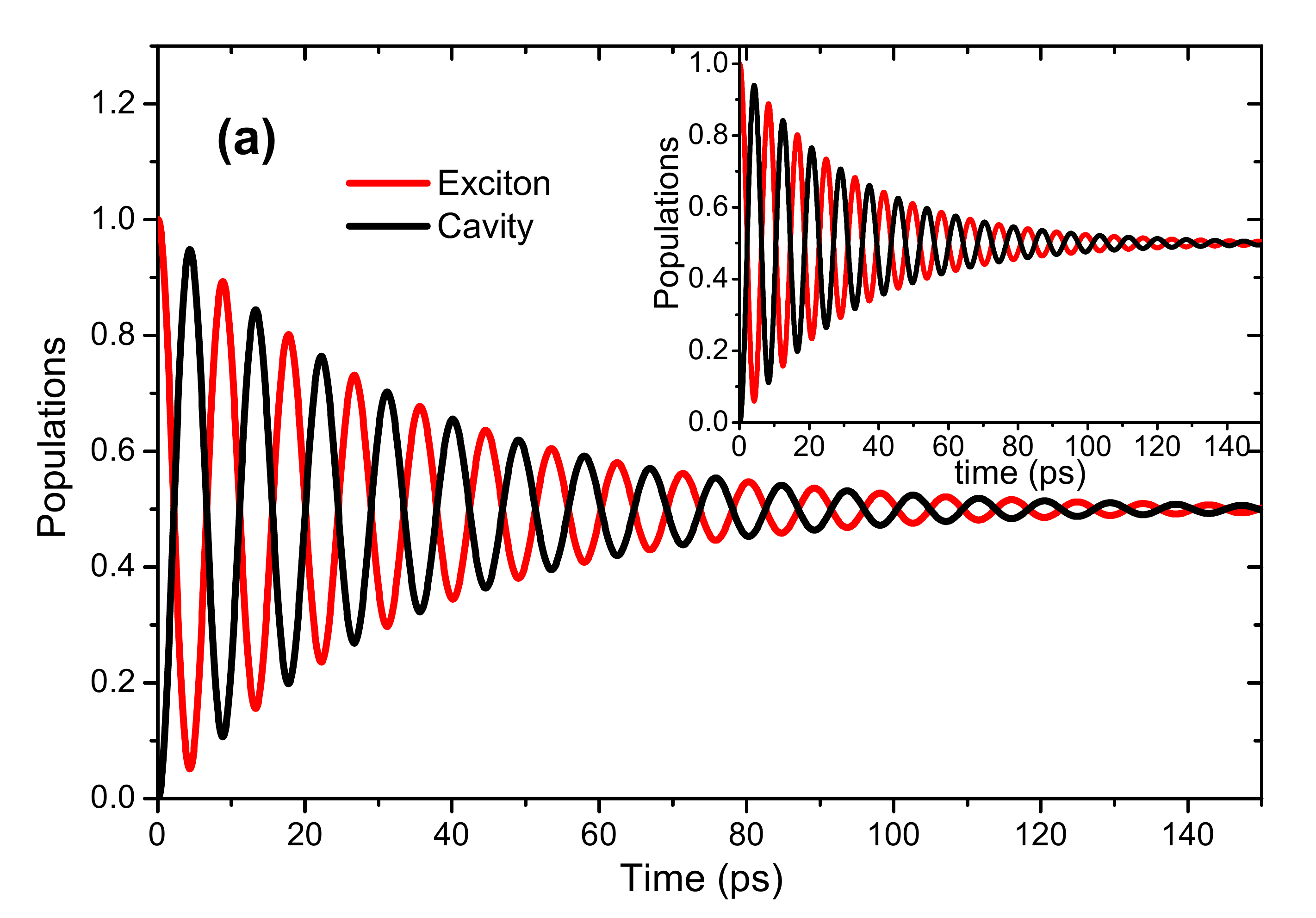}
\includegraphics[width=0.45\textwidth]{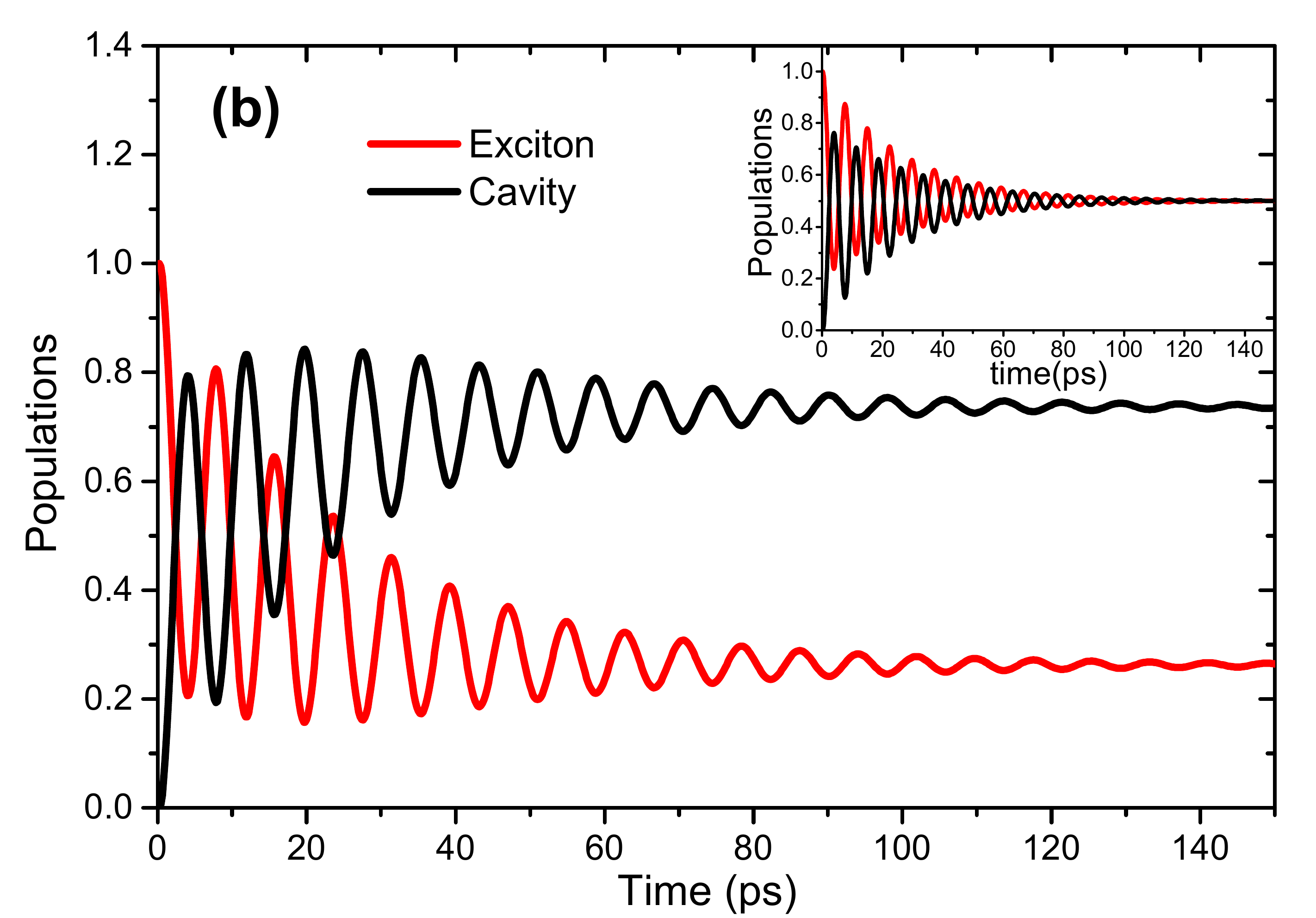}  
\includegraphics[width=0.45\textwidth]{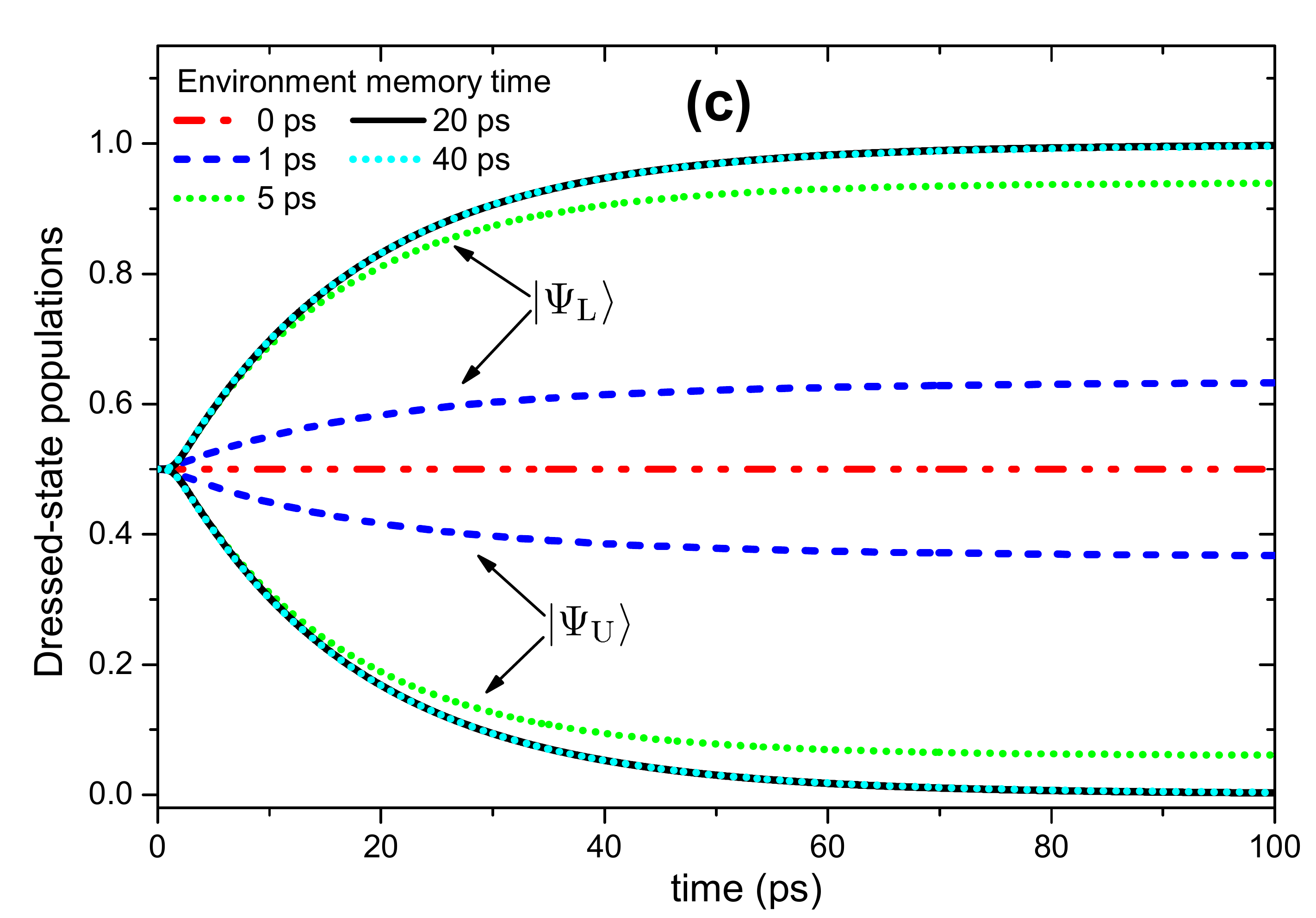} 
\includegraphics[width=0.45\textwidth]{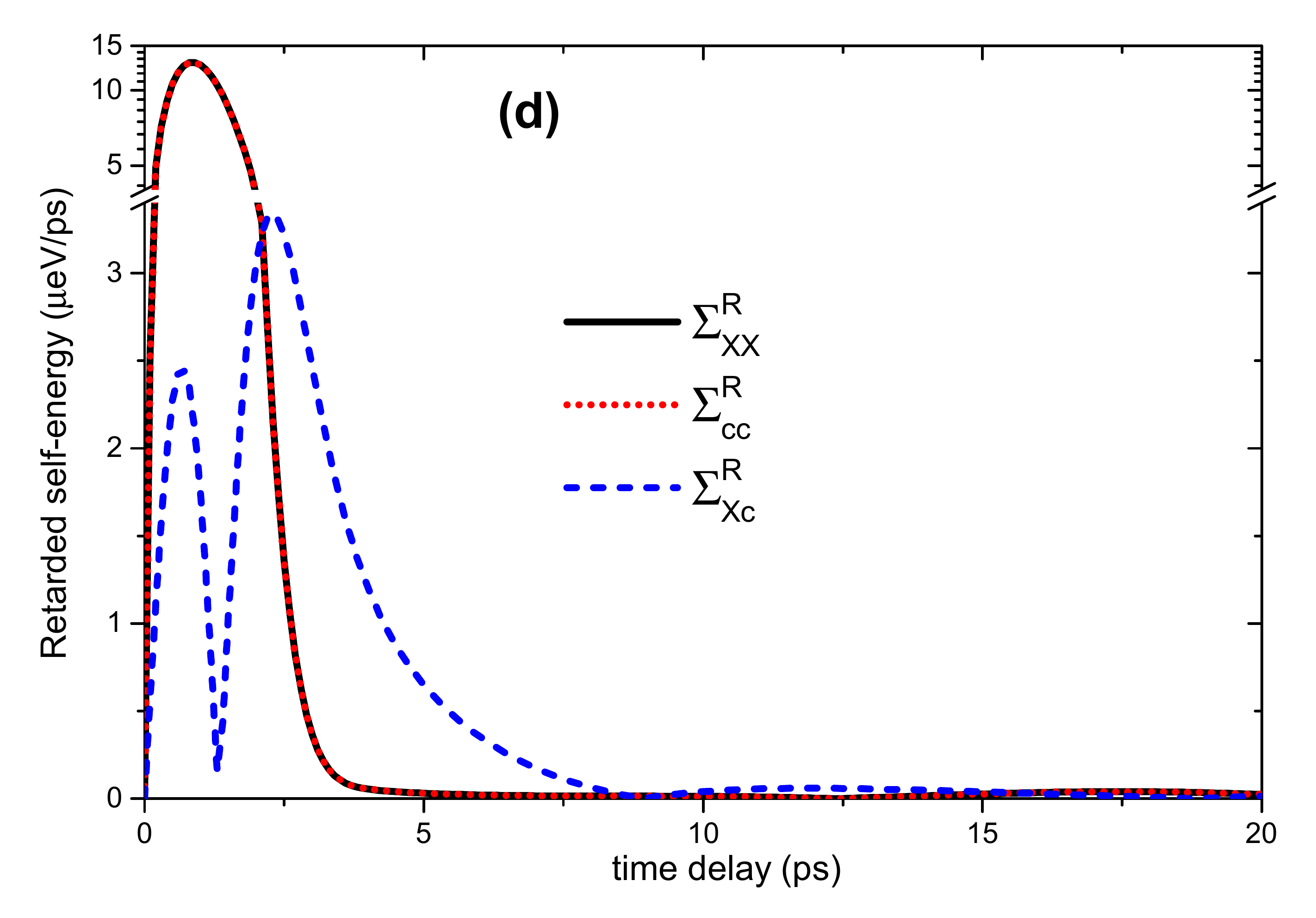} 
\end{centering}
\caption{(Color online). Dynamics of a strongly-coupled QD-cavity system in presence of a zero-temperature phonon bath. (a-b): The populations of $|\Psi_{\text{X}}\rangle=|e,0\rangle$ (exciton) and $|\Psi_{\text{c}}\rangle=|g,1\rangle$ (cavity) are plotted as a function of time following an initial excitation of the QD. The exciton-cavity detuning in is set to (a) $\tilde{\omega}_{\text{X}}-\omega_{\text{c}} = 0$ and (b) $\tilde{\omega}_{\text{X}}-\omega_{\text{c}} = $ 250~\micro eV in (b).
The insets in (a) and (b) show the population dynamics assuming a Markovian pure dephasing term $\gamma^* = 80$~\micro eV instead of the non-Markovian phonon bath.
(c) Populations of the dressed states $|\psi_ {\text{U/L}}\rangle = (|\Psi_{\text{X}}\rangle \mp |\Psi_{\text{c}}\rangle)/\sqrt{2}$ in the zero-detuning case ($\tilde{\omega}_{\text{X}}-\omega_{\text{c}} = 0$) as a function of time for various maximum correlation time $\tau_{\text{c}}^{\text{m}}$ for the phonon bath.
(c) Retarded self-energy matrix elements $\Sigma_{\text{XX}}(t+\tau,t)$ (black full line), $\Sigma_{\text{cc}}(t+\tau,t)$ (red dotted line), $\Sigma_{\text{Xc}}(t+\tau,t)$ (blue dashed line) as a function of the time delay $\tau$ for $\tilde{\omega}_{\text{X}}-\omega_{\text{c}} = 0$ ($\tau_{\text{c}}^{\text{m}}$ is set to 40~ps).
The parameters are: $\hbar g=250~$\micro eV , $D=20$~eV, $\sigma = 5$~nm, $\kappa = 0$, $\gamma=0$, $\gamma^*=0$ (apart from insets) and $T= 0$.
}
\label{Rabi}
\end{figure*}

The model is applied to the study the dynamics from the initial state $|e,0\rangle$, i.e. after an instantaneous excitation of the QD exciton, the photonic cavity being initially empty.
The possible occupied states in the subsequent dynamics are only $|e,0\rangle$, $|g,1\rangle$ and $|g,0\rangle$.
As the decay to $|g,0\rangle$ is incoherent and irreversible, it is sufficient to consider the nonequilibrium Green's functions in the two-level basis ($|\Psi_{\text{X}}\rangle=|e,0\rangle$,$|\Psi_{\text{c}}\rangle=|g,1\rangle$). 
The population of the $|g,0\rangle$ state is then deduced by population conservation.
In the following, we study the system dynamics, the photon spectra and the single-photon indistinguishability. Our approach is compared with various analytical expectations and an exact-diagonalization approach.

\subsection{Time dynamics and non-Markovian effects}

We first study the QD-cavity dynamics in the strong-coupling regime.
 For the sake of clarity, the Markovian dissipative terms are set to zero ($\gamma=\gamma^* = \kappa=0$), so that the system dynamics reduces to a trace-conserving two-level system consisting of an exciton state ($|\Psi_{\text{X}}\rangle=|e,0\rangle$) and a cavity state ($|\Psi_{\text{c}}\rangle=|g,1\rangle$).
The temperature is set to 0K.
The time evolution of the exciton and cavity  populations are reported in Fig.~\ref{Rabi}~(a) and (b) for QD-cavity detuning $(\tilde{\omega}_{\text{X}}-\omega_{\text{c}})$ of respectively 0 and 250~\micro eV.
To compare qualitatively with a Markovian behavior, the insets show the population dynamics when no phonon coupling is considered but instead a pure dephasing term of $\gamma^* = 80$~\micro eV.
In the case of a finite detuning (Fig.~\ref{Rabi}b), the dynamics clearly differs from the Markovian case.

To get more insights, the dressed polariton states  $|\psi_ {\text{U/L}}\rangle = \alpha_ {\text{U/L}} |\Psi_{\text{X}}\rangle + \beta_ {\text{U/L}}|\Psi_{\text{c}}\rangle)$, i.e. the eigenstates of the exciton-cavity system, are considered \footnote{Note that as we are considering a polaron basis, the computed polariton states involve a displaced lattice equilibrium in the exciton state.}.
At 0K, the system is expected to relax towards the lower polariton level $|\psi_ {\text{L}}\rangle$.
In the case of equal detuning and coupling strength ($\hbar g = \hbar \delta = 250$~\micro eV), this can be checked on Fig.~\ref{Rabi}(b): in the long time limit, the populations are in perfect agreement with the analytical expectation
 for the lower polariton state $|\beta_ {\text{L}}|^2 = 2/(5-\sqrt{5})\simeq 0.72$
 \footnote{This analytical expression for the ground state population is verified for any two-level system ($|0\rangle,|1\rangle$) with a Hamiltonian of the form $H= \delta |1\rangle \langle 1| + \delta(|0\rangle \langle 1|+|1\rangle \langle 0|)$.}.
In the zero-detuning case, the populations of the polaritons states $|\psi_ {\text{U/L}}\rangle$ are plotted on Fig.~\ref{Rabi}(c). As expected, the full calculation (corresponding here to $\tau_{\text{c}}^{\text{m}}=$~20 or 40~ps) shows a relaxation towards the lower polariton state $|\psi_ {\text{L}}\rangle$.
Interestingly, the full calculation  is compared to calculations in which the maximum memory time $\tau_{\text{c}}^{\text{m}}$ is numerically bounded to smaller values (5, 1 and 0~ps). For such shorter correlation time of the phonon reservoir, the system does not relax towards its ground state.
This behavior can be understood in the following way: in the short memory time limit, the energy splitting between the two polaritons is not resolved by the bath and the system do not thermalize. This corresponds to the Markovian limit where the polariton populations are not affected by the coupling to the bath. On the other hand, thermal relaxation takes place as soon as the memory time used for the computation overcomes the physical bath's memory time (Fig.~\ref{Rabi}(d)). 
In the full calculation (i.e. for sufficiently long correlation time), we have checked that for various QD-cavity detunings, QD-cavity coupling strengths and temperatures the system always relax towards thermal equilibrium (not shown).
We stress that such relaxation of the system towards the thermal equilibrium for all detunings and coupling strengths cannot be described with a fixed Markovian dissipative term in the Lindblad master equation, since, as stated above, a Markovian bath does not gain information about the open system energy.
Note that Markovian models for phonon environment that have been used previously \cite{hohenester2010cavity,majumdar2011phonon,muller2015ultrafast} make use of dissipative terms that depends explicitly on the polariton energy level positions.

The matrix elements of the retarded self-energy $\Sigma^{\text{R}}(t,t+\tau)$  are plotted on Fig.~\ref{Rabi}d as a function of $\tau$
in the zero-detuning case. It confirms that memory effects from the phonon bath back on the QD-cavity system last on the few-ps time scale. This time-scale can be interpreted as the time taken for the emitted acoustic phonons to escape from the exciton wavefunction region.
This phonon memory effect is further evidenced by the short time dynamics on Fig.~\ref{Rabi}(c): on the picosecond time scale after the excitation, the population relaxation of the dressed states is strongly non-exponential. Indeed, for delays smaller than the phonon escape time, phonon scattering processes are reversible and no relaxation between the polariton states is observed.

\begin{figure}
\begin{centering}
\includegraphics[width=0.48\textwidth]{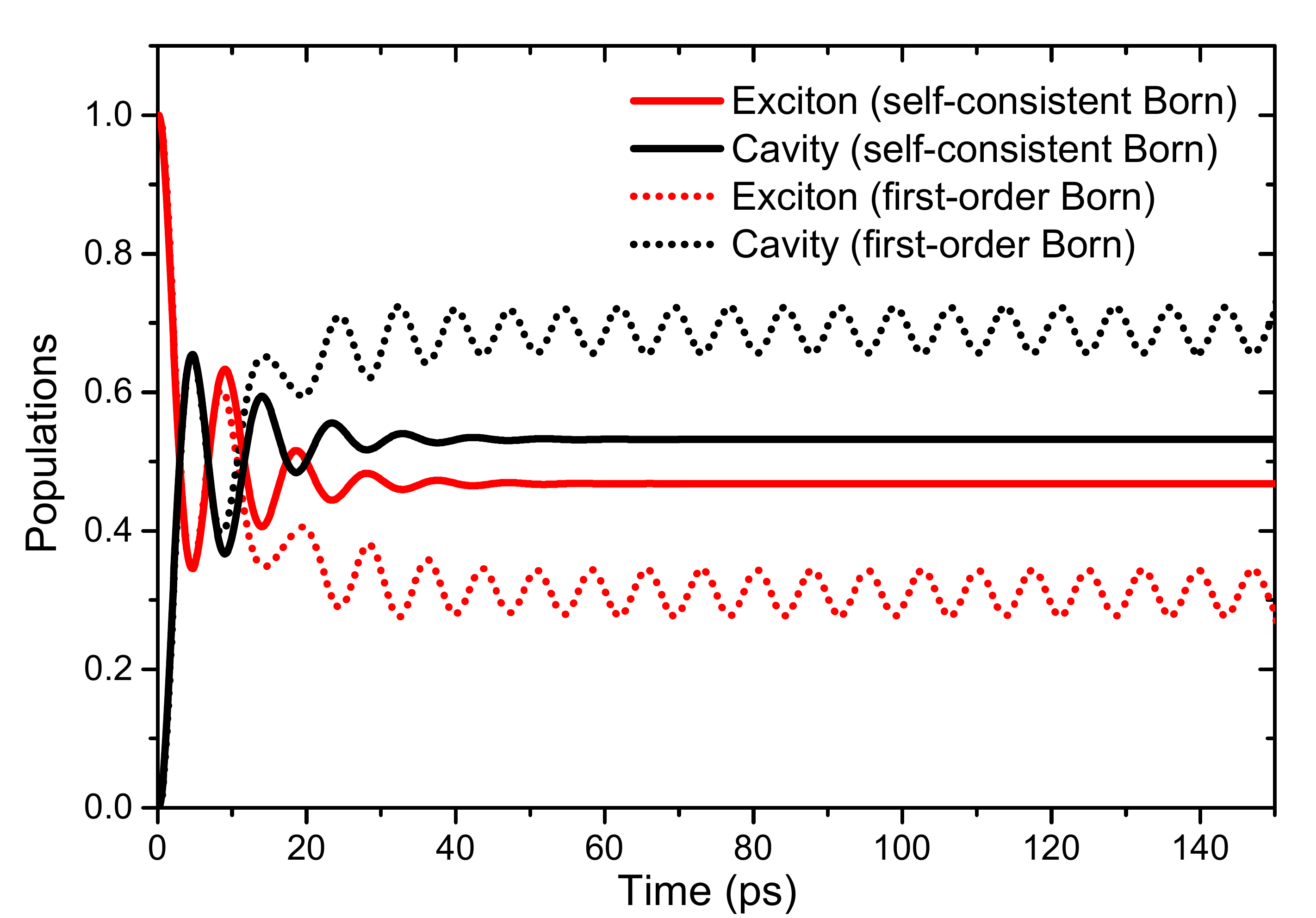} 
\end{centering}
\caption{(Color online). 
Population dynamics of a strongly-coupled QD-cavity system with the same parameters as in Fig.~\ref{Rabi}b except than the lattice temperature is set to $T=20$~K.
The full calculation using the self-consistent born approximation (full lines) is compared with the single-Born approximation (dotted lines).
}
\label{Born}
\end{figure}

Fig.~\ref{Born} shows the population dynamics with the same parameters as in Fig.~\ref{Rabi}b but at a finite temperature of 20~K.
The relaxation towards the thermal equilibrium is observed in the full calculation.
This full calculation is compared with the first-order Born approximation, in which the self-energies (Eqs.~\ref{sigmalesser}, \ref{sigmaretarded}) are calculated from a non-interacting GF $\hat{G}_0$ (solution of $\hat{H}_{\text{int}}=0$) instead of the actual GF $\hat{G}$ used in the self-consistent equations.
This first-order Born approximation (which is of second-order with respect to the electron-phonon interaction) gives a large discrepancy with respect to the self-consistent Born approximation.
Indeed, the relaxation dynamics involves several scattering processes at finite temperature, 
while the first-order Born approximation accounts for a single scattering event.
In the long-time limit, an infinite number of scattering events has to be describe, which imposes to go beyond
finite-order perturbation theories \cite{goan10,kaer2014decoherence} or truncated Hilbert space treatment for the phonon reservoir \cite{kaer2013microscopic}.



\subsection{Cavity-radiated spectra}

\begin{figure}
\begin{centering}
\includegraphics[width=0.48\textwidth]{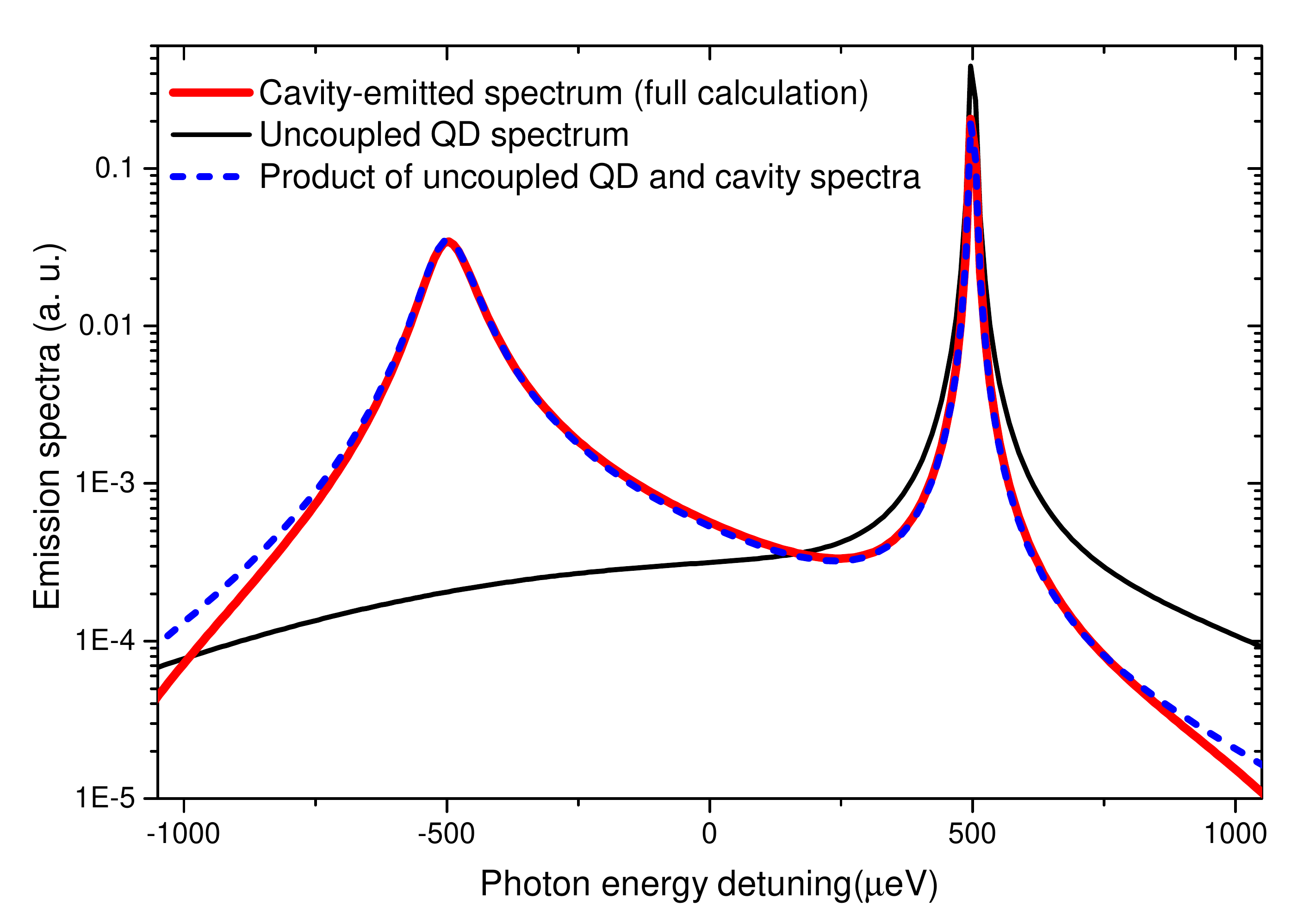} 
\end{centering}
\caption{(Color online). Photon emission spectrum from a cavity fed by a higher energy QD in the weak coupling limit (red thick line). The thin black line indicates the bare QD emission spectrum. The dotted blue line is the product of the bare QD emission by the free cavity one, which is expected to coincide with the cavity emission in the weak coupling limit. The parameters are $\hbar g=1~$\micro eV, $\tilde{\omega}_{\text{X}}-\omega_{\text{c}} =1000~$\micro eV, $\hbar \kappa = 100~$\micro eV, $\hbar \gamma=10~$\micro eV, $\gamma^*=0$, $T= 4$~K, $D=9.8$~eV, $\sigma = 5$~nm.
}
\label{CavityFeeding}
\end{figure}

We now consider non-zero decay rates for the cavity ($\kappa$) and the QD ($\gamma$).
The spectrum of the photons emitted by the cavity reads
\begin{equation}
S(\omega)=  \int_{0}^{\infty} dt \int_0^{\infty} d\tau e^{-i\omega \tau}\langle \hat{a}^{\dagger}(t+\tau)\hat{a}(t)\rangle
\end{equation}
The involved correlator is given by the cavity component of the lesser GF
\begin{equation}
\langle \hat{a}^{\dagger}(t+\tau)\hat{a}(t)\rangle = -iG_{\text{c},\text{c}}(t+\tau,t)
\end{equation}
where the c index represents the one-photon cavity state $|g,1\rangle$, as only a single excitation is considered.

In the limit of a small coupling with respect to the difference between cavity and emitter linewidths ($g \ll |\kappa-\gamma|)$, it has been shown that the spectrum emitted by the cavity can be approximated by the product of the bare QD polaron spectrum multiplied by the bare cavity one\cite{Portalupi2015Bright}:
\begin{equation}
S_{\text{weak-coupling}}(\omega) \propto S_{\text{polaron}}(\omega)  \times S_{\text{cavity}}(\omega) .
\end{equation}
To assess the validity of our model in this limiting case, we compare the calculated spectrum in a weak coupling regime with the product of the cavity spectrum by the QD spectrum on Fig.~\ref{CavityFeeding}. A very good agreement is found, demonstrating the accuracy of our model in the weak-coupling limit.


\subsection{Photon indistinguishability}

\begin{figure}
\begin{centering}
\includegraphics[width=0.48\textwidth]{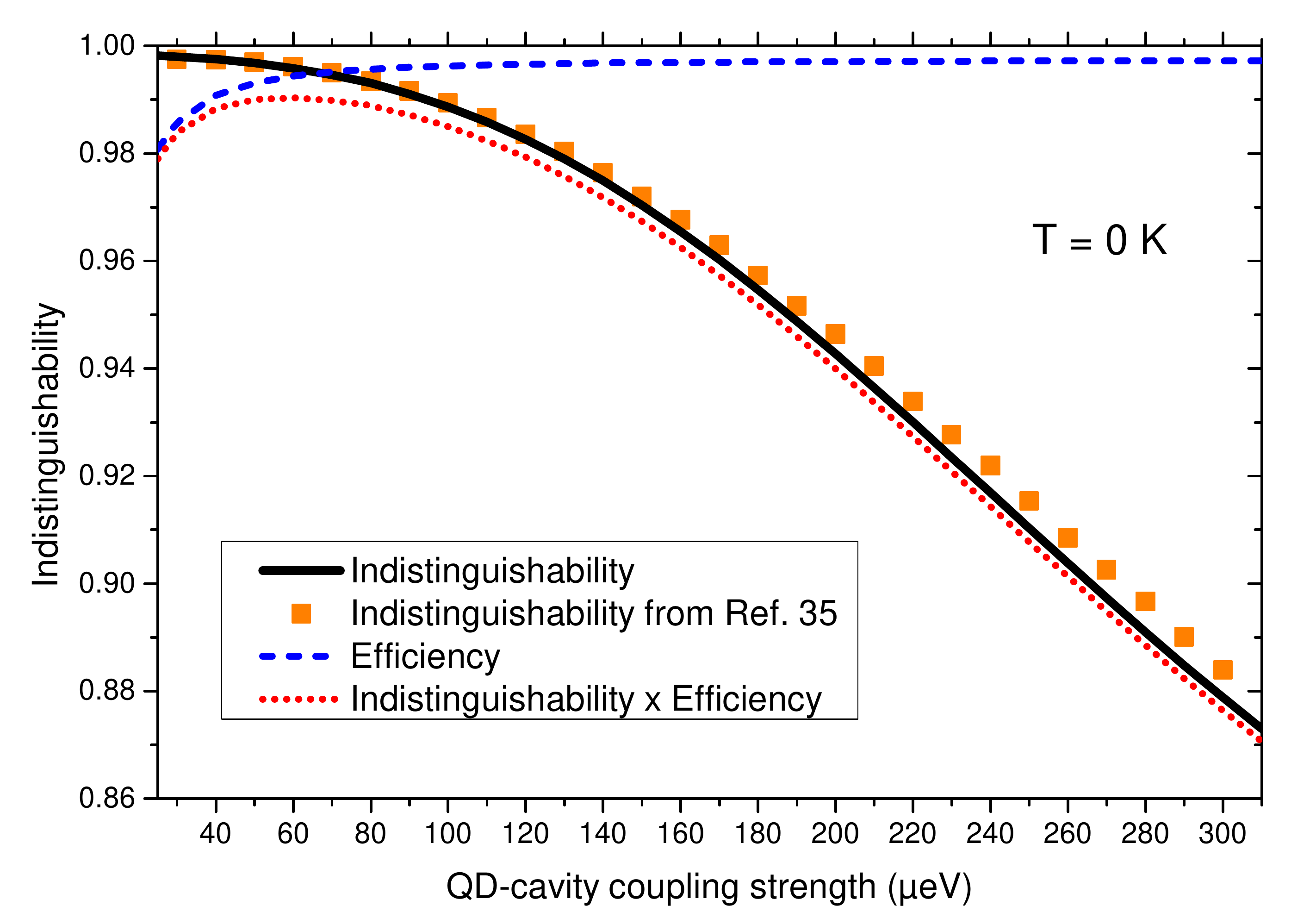}
\includegraphics[width=0.48\textwidth]{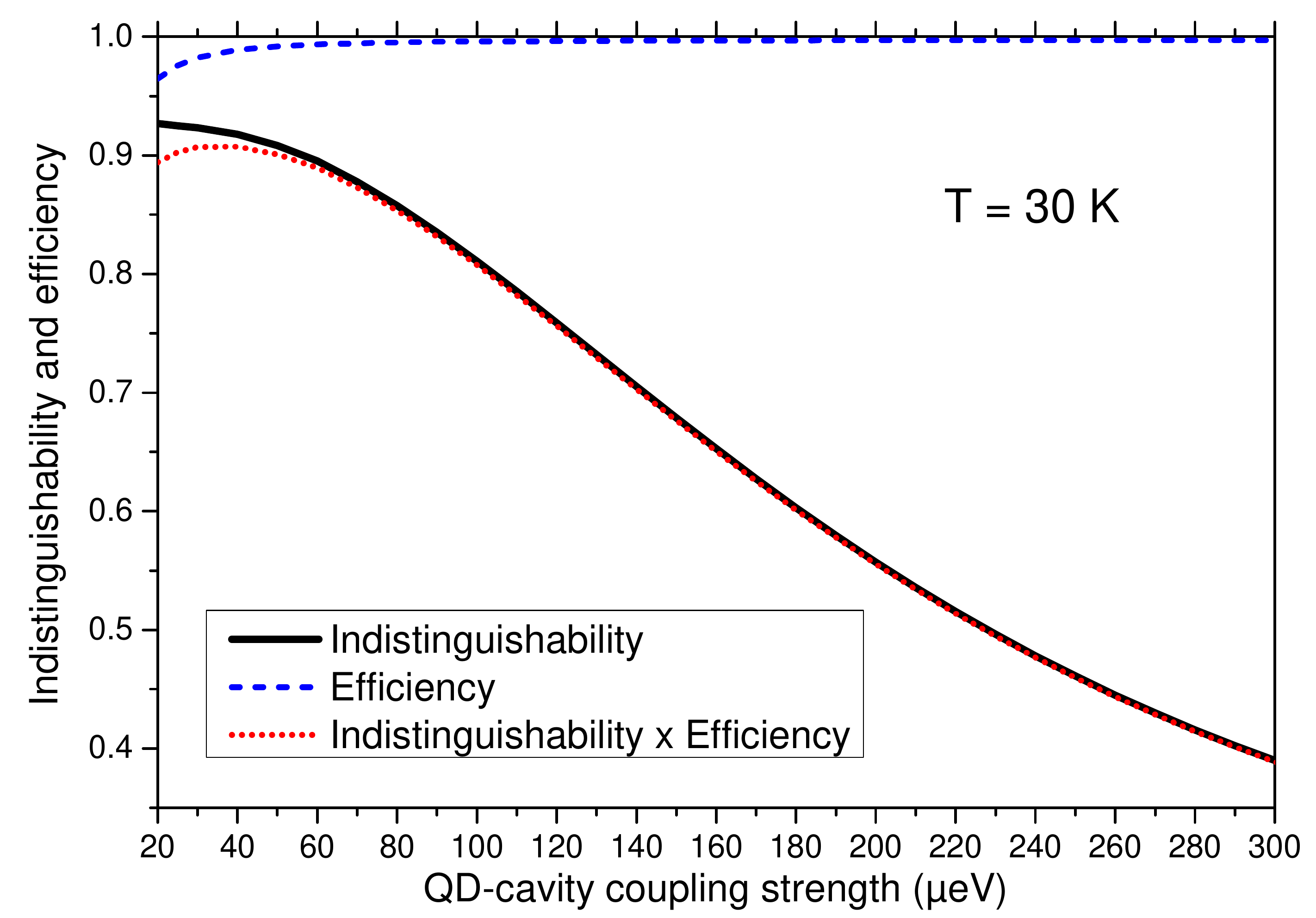}
\end{centering}
\caption{(Color online). Photon indistinguishability as a function of the QD-cavity coupling strength for lattice temperatures of 0K (upper panel) and 30K (lower panel). The other parameters are the same as in Ref.~\onlinecite{kaer2013microscopic}: $\hbar \kappa = 125~$\micro eV, $\gamma=0.5~$ns$^{-1}$, $\gamma^*=0$, $D=9.8$~eV, $\sigma = 5$~nm.}
\label{I-vs-g}
\end{figure}

The degree of indistinguishability of the single photons emitted by the cavity is an important characteristic in view of applications of single photons in quantum computation\cite{knill2001scheme}.
The indistinguishability figure of merit describes the probability for two single-photon wavepackets to interfere coherently when being sent simultaneously on a beam splitter.
This indistinguishability figure of merit for the photons emitted by the cavity reads \cite{grange2015cavity}:
\begin{equation}
I=  \frac{\int_{0}^{\infty} dt \int_0^{\infty} d\tau \vert\langle \hat{a}^{\dagger}(t+\tau)\hat{a}(t)\rangle\vert^2}{\int_{0}^{\infty} dt \int_0^{\infty} d \tau \langle \hat{a}^{\dagger}(t)\hat{a}(t)\rangle\langle \hat{a}^{\dagger}(t+\tau)\hat{a}(t+\tau)\rangle} .
\end{equation}
In Fig.~\ref{I-vs-g} we report calculations of indistinguishability of the photons emitted by the cavity as a function of the QD-cavity coupling strength at temperatures of 0~K and 30~K.
At 0K, the indistinguishability is found to be in very good agreement the exact diagonalization approach reported in Ref.~\onlinecite{kaer2013microscopic}.
In addition we report calculations at finite temperature (30K) which has not been tractable using exact diagonalization.
For any temperature, as the exciton-cavity coupling strength increases, the indistinguishability decreases. This is due to an increasing density of states of acoustic phonons available to induce scattering between the two polariton states.
The efficiency, defined as the probability to emit a photon from the cavity mode for an initial excitation of the QD, is also shown in Fig.~\ref{I-vs-g}. Its expression reads:
\begin{equation}
\beta= \kappa \int_{0}^{\infty}  \langle \hat{a}^{\dagger}(t)\hat{a}(t)\rangle.
\label{Betaformula}
\end{equation}
In the weak coupling regime, the efficiency increases with increasing coupling strength, and saturates in the strong coupling regime at $\kappa / (\kappa+ \gamma)$.
The product $\beta I$ is also plotted in Fig.~\ref{I-vs-g}, showing that there is a necessary trade-off between indistinguishability and efficiency when tuning the coupling strength.

\begin{figure}
\begin{centering}
\includegraphics[width=0.48\textwidth]{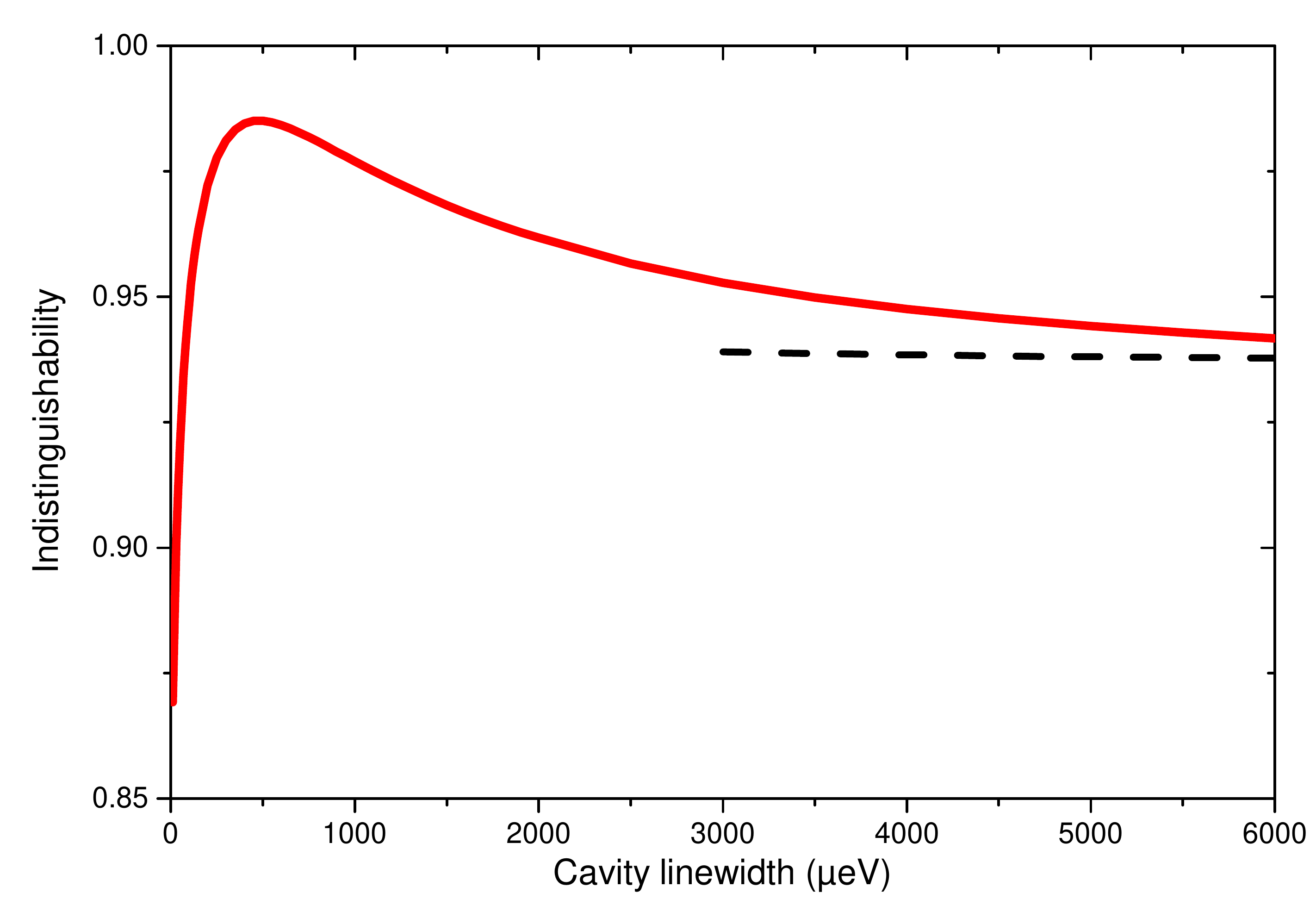} 
\end{centering}
\caption{(Color online). Photon indistinguishability as a function of the cavity linewidth $\hbar \kappa$ in solid line. The dashed line indicates $I_{\kappa \rightarrow \infty}$, i.e. the expected limit when the cavity linewidth becomes large compared to the phonon energies.
The fixed parameters are $\hbar g= 200~$\micro eV, $\gamma=50~$ns$^{-1}$, $\gamma^*=0$,  $D=9.8$~eV, $\sigma = 5$~nm.}
\label{I-vs-kappa-limit}
\end{figure}

Fig.~\ref{I-vs-kappa-limit} shows a calculation of photon indistinguishability as a function of the cavity linewidth $\hbar \kappa$.
For low cavity linewidths $\kappa < 2 g$, the system is in the strong exciton-photon coupling regime. First, the indistinguishability increases with increasing cavity linewidth. Indeed, as the system enters the weak coupling regime, the phonon-assisted scattering effect between the splitted polariton states (as discussed above) tends to disappear. Beyond, a further increase in the cavity linewidth decreases the indistinguishability, as a large cavity linewidth enables the higher energy phonons to assist the photon emission as well.

We focus below on the large cavity linewidth limit, where an analytical expression exists.
Indeed, in the limit of cavity linewidths much larger than (i) the coupling strength ($\kappa \gg g$) and (ii) the phonon continuum effectively coupled to the QD ($\kappa \gg \hbar c_s / \sigma$), 
the cavity can be adiabatically eliminated as it decays faster than the QD coherent evolution~\cite{kaer2013role,grange2015cavity}. Hence the cavity coherences follows directly the QD coherences
$\langle \hat{a}^{\dagger}(t+\tau)\hat{a}(t)\rangle \propto \langle \hat{c}^{\dagger}(t+\tau)\hat{c}(t)\rangle$.
This leads to the following analytical results\cite{kaer2013role}:
\begin{equation}
I_{\kappa \rightarrow \infty} = \Gamma \int d\tau \exp[-\Gamma \tau + 2\text{Re}[\phi(\tau) - \phi(0)]]  
\end{equation}
where $ \Gamma = \gamma + 4g^2/\kappa$ is the effective decay rate of the exciton.
This limit is indicated as a dashed line in Fig.~\ref{I-vs-kappa-limit}.
The calculated indistinguishability is found to converge slowly towards this limit.
This is in strong contrast with the non-Markovian second-order perturbation theory approach recently reported by Kaer and M\o rk \cite{kaer2014decoherence}, which has been shown to break down at large cavity linewidth.
The slow convergence towards the large-cavity limit demonstrates the need of an accurate modeling if the cavity linewidth is not an order of magnitude larger than the phonon sidebands energy scale.

\section{Summary and outlook}

We have presented a nonequilibrium Green's function approach to calculate the dynamics of coupled QD-cavity systems interacting with a phonon bath.
We use the self-consistent Born approximation within a polaron basis, which accounts for an infinite number of electron-phonon-photon scattering processes. 
Calculations of dynamics, photon spectra and photon indistinguishability are reported.
Our approach is shown to be in good agreement with various exact solutions of the problem in some limiting cases, yet allowing to tackle this problem for a 
unprecedented range of temperatures and cavity-QD coupling strengths.

Our model can be extended to time-dependent Hamiltonian. This will enable the study of phonon-induced damping of Rabi rotations in quantum dots \cite{machnikowski2004resonant,vagov2007nonmonotonic,ramsay2010damping,mccutcheon2010quantum,glassl2011influence,debnath2012chirped,manson2016polaron,mccutcheon2016optical,nazir2016modelling} and to account for resonant excitation used in indistinguishable photon generation \cite{somaschi2015near,giesz2016coherent}.
Our method can be applied to other solid-state cavity-QED systems such as nitrogen-vacancy centers in diamond \cite{johnson2015tunable}, and more generally to other open quantum systems coupled to non-Markovian environment.

\section*{Acknowledgements}
This work has been funded by the European Union's Seventh Framework Programme (FP7) under Grant Agreement No. 618078 (WASPS). We thank P. Kaer for providing the data of Ref.~\onlinecite{kaer2013microscopic}.

\appendix

\section{Full-Hilbert-space nonequilibrium Green's functions}

The NEGF formalism is standardly used for calculating single-particle GFs of either fermions or bosons within many-particle systems.
The dimensions of the GFs are hence typically much smaller than the size of the full Hilbert space.
Here instead we define GFs within the full Hilbert space of a small quantum system interacting with an external bath. The motivation is to extend the density matrix formalism to two-times correlation functions in a manner such as the Kadanoff-Baym equations of the NEGF formalism can be used. 


Within many-body systems, the single-particle fermionic GFs are standardly defined \cite{haug2008quantum}. This standard definition applied directly to the fermionic creation/annihilation operators of the full Hilbert space ($ \hat{\Psi}_i^{(\dagger)}$) would lead to:
\begin{subequations}
\begin{equation}
g_{i,j}^<(t_1,t_2) = i \langle \hat{\Psi}^{\dagger}_j (t_2) \hat{\Psi}_i(t_1) \rangle
\end{equation}
\begin{equation}
g_{i,j}^>(t_1,t_2) = -i \langle \hat{\Psi}_i(t_1) \hat{\Psi}^{\dagger}_j (t_2)  \rangle
\end{equation}
\begin{equation}
\hat{g}^{R}(t_1,t_2) = \Theta(t_1-t_2)\left[ \hat{g}^>(t_1,t_2)-\hat{g}^<(t_1,t_2) \right] ,
\end{equation}
\begin{equation}
\hat{g}^{A}(t_1,t_2) = -\Theta(t_2-t_1)\left[ \hat{g}^>(t_1,t_2)-\hat{g}^<(t_1,t_2) \right] ,
\end{equation}
\label{standardNEGF}
\end{subequations}
In this many-body standard formulation, the retarded and advanced Green's functions depend on the nonequilibrium state described by the lesser and greater GFs, as they describe propagation of excitations with respect to the nonequilibrium state. If one applies this definition directly to the case of a full Hilbert space, it would result in an unphysical nonlinear dependence of the density matrix with respect to the initial conditions.
The reason is that the above definition is meant to prevent scattering of a fermion towards an already occupied level, i.e. imposing the Pauli exclusion principle.
In contrast, in the dynamics of a quantum state within its full Hilbert space, such exclusion principle must be not included.

To suppress this nonlinearity in the equations of motion arising from the exclusion principle, in the following we renormalize the GFs of the standard formalism.
If we renormalize the initial density matrix $\rho(0)$ by a constant $\epsilon$, the initial renormalized lesser GFs reads
\begin{equation}
\hat{g}_{\epsilon}^<(0,0) = i \epsilon \hat{\rho} (0).
\end{equation}
We then consider the GFs $\hat{g}_{\epsilon}(t_1,t_2)$ that are solutions of the Kadanoff-Baym equations with the above initial condition for the density matrix at $t=0$. 
As discussed above, the Eqs.~\ref{standardNEGF} are not expected to correctly describe the system dynamics for latter times ($t_1$, $t_2$).
Instead we assume that the renormalized GF $\hat{g}_{\epsilon}$ correctly accounts for the system dynamics in the small $\epsilon$ limit.
In the limit of small $\epsilon$, the lesser GF becomes small compared to the greater GF, so that the retarded GF reads:
\begin{equation}
\hat{g}_{\epsilon}^{R}(t_1,t_2) \underset{\epsilon \to 0}{=} -i \Theta(t_1-t_2) \hat{g}_{\epsilon}^>(t_1,t_2)
\end{equation}
As the excitation probability is vanishingly small, the greater GF (and hence the retarded GF) can then be equivalently evaluated on a vacuum state

\begin{equation}
\lim_{\epsilon \to 0} \hat{g}_{(ij) \epsilon}^>(t_1,t_2) =  -i\langle \hat{\Psi}_i(t_1) \hat{\Psi}^{\dagger}_j (t_2)  \rangle_{\text{vac}} ,
\end{equation}
where vac stands for a fictitious vacuum state (i.e. not part of the physical Hilbert space).
Another consequence, is that the retarded self-energies $\hat{\sigma}^R_{\epsilon}$, which depends in principle on both $\hat{g}^<_{\epsilon}$ and $\hat{g}^R_{\epsilon}$, depends only on $\hat{g}^R_{\epsilon}$.
Hence in this limit of small $\epsilon$ the Kadanoff-Baym equations for $\hat{g}_{\epsilon}$ give the expected linear regime for the lesser GF $\hat{g}^<_{\epsilon}$ with respect to the initial density matrix $\hat{\rho} (0)$.
As a consequence we use these renormalized nonequilibrium GFs for the full Hilbert space dynamics
\begin{equation}
\hat{G}^<(t_1,t_2) = \lim_{\epsilon \to 0} \frac{\hat{g}_{\epsilon}^<(t_1,t_2)}{\epsilon}
\end{equation}
\begin{equation}
\hat{G}^{R}(t_1,t_2) = \lim_{\epsilon \to 0} \hat{g}_{\epsilon}^{R}(t_1,t_2)
\end{equation}
Owing to the expected linearity of the lesser GFs $G_{i,j}^<(t_1,t_2)$ with respect to the initial condition, these GFs for the full Hilbert space can be redefined directly as:
\begin{equation}
G_{i,j}^<(t_1,t_2) = i \langle \hat{\Psi}^{\dagger}_j (t_2) \hat{\Psi}_i(t_1) \rangle
\end{equation}
\begin{equation}
G_{i,j}^{R}(t_1,t_2) = -i\Theta(t_1-t_2)\langle \hat{\Psi}_i(t_1) \hat{\Psi}^{\dagger}_j (t_2)  \rangle_{\text{vac}}.
\end{equation}
The above definitions are used in the present work. Note that this retarded GF can also be expressed as $\hat{G}^{R}(t_1,t_2) = -i\Theta(t_1-t_2) \hat{U}(t_1,t_2)$ where $\hat{U}(t_1,t_2)$ is the evolution operator corresponding to the total Hamiltonian.
The self-energies can then be calculated directly following the above definitions, and the Kadanoff-Baym equations applied to calculate the dynamics.

\bibliographystyle{apsrev}
\bibliography{C:/Users/thomas.grange/GoogleDrive/papers/biblio/biblio}

\begin{thebibliography}{55}
\expandafter\ifx\csname natexlab\endcsname\relax\def\natexlab#1{#1}\fi
\expandafter\ifx\csname bibnamefont\endcsname\relax
  \def\bibnamefont#1{#1}\fi
\expandafter\ifx\csname bibfnamefont\endcsname\relax
  \def\bibfnamefont#1{#1}\fi
\expandafter\ifx\csname citenamefont\endcsname\relax
  \def\citenamefont#1{#1}\fi
\expandafter\ifx\csname url\endcsname\relax
  \def\url#1{\texttt{#1}}\fi
\expandafter\ifx\csname urlprefix\endcsname\relax\def\urlprefix{URL }\fi
\providecommand{\bibinfo}[2]{#2}
\providecommand{\eprint}[2][]{\url{#2}}

\bibitem[{\citenamefont{O'Brien et~al.}(2009)\citenamefont{O'Brien, Furusawa,
  and Vu{\v{c}}kovi{\'c}}}]{o2009photonic}
\bibinfo{author}{\bibfnamefont{J.~L.} \bibnamefont{O'Brien}},
  \bibinfo{author}{\bibfnamefont{A.}~\bibnamefont{Furusawa}}, \bibnamefont{and}
  \bibinfo{author}{\bibfnamefont{J.}~\bibnamefont{Vu{\v{c}}kovi{\'c}}},
  \bibinfo{journal}{Nature Photonics} \textbf{\bibinfo{volume}{3}},
  \bibinfo{pages}{687} (\bibinfo{year}{2009}).

\bibitem[{\citenamefont{Hennessy et~al.}(2007)\citenamefont{Hennessy, Badolato,
  Winger, Gerace, Atat{\"u}re, Gulde, F{\"a}lt, Hu, and
  Imamo{\u{g}}lu}}]{hennessy2007quantum}
\bibinfo{author}{\bibfnamefont{K.}~\bibnamefont{Hennessy}},
  \bibinfo{author}{\bibfnamefont{A.}~\bibnamefont{Badolato}},
  \bibinfo{author}{\bibfnamefont{M.}~\bibnamefont{Winger}},
  \bibinfo{author}{\bibfnamefont{D.}~\bibnamefont{Gerace}},
  \bibinfo{author}{\bibfnamefont{M.}~\bibnamefont{Atat{\"u}re}},
  \bibinfo{author}{\bibfnamefont{S.}~\bibnamefont{Gulde}},
  \bibinfo{author}{\bibfnamefont{S.}~\bibnamefont{F{\"a}lt}},
  \bibinfo{author}{\bibfnamefont{E.~L.} \bibnamefont{Hu}}, \bibnamefont{and}
  \bibinfo{author}{\bibfnamefont{A.}~\bibnamefont{Imamo{\u{g}}lu}},
  \bibinfo{journal}{Nature} \textbf{\bibinfo{volume}{445}},
  \bibinfo{pages}{896} (\bibinfo{year}{2007}).

\bibitem[{\citenamefont{Englund et~al.}(2007)\citenamefont{Englund, Faraon,
  Fushman, Stoltz, Petroff, and Vu{\v{c}}kovi{\'c}}}]{englund2007controlling}
\bibinfo{author}{\bibfnamefont{D.}~\bibnamefont{Englund}},
  \bibinfo{author}{\bibfnamefont{A.}~\bibnamefont{Faraon}},
  \bibinfo{author}{\bibfnamefont{I.}~\bibnamefont{Fushman}},
  \bibinfo{author}{\bibfnamefont{N.}~\bibnamefont{Stoltz}},
  \bibinfo{author}{\bibfnamefont{P.}~\bibnamefont{Petroff}}, \bibnamefont{and}
  \bibinfo{author}{\bibfnamefont{J.}~\bibnamefont{Vu{\v{c}}kovi{\'c}}},
  \bibinfo{journal}{Nature} \textbf{\bibinfo{volume}{450}},
  \bibinfo{pages}{857} (\bibinfo{year}{2007}).

\bibitem[{\citenamefont{Kasprzak et~al.}(2010)\citenamefont{Kasprzak,
  Reitzenstein, Muljarov, Kistner, Schneider, Strauss, H{\"o}fling, Forchel,
  and Langbein}}]{kasprzak2010up}
\bibinfo{author}{\bibfnamefont{J.}~\bibnamefont{Kasprzak}},
  \bibinfo{author}{\bibfnamefont{S.}~\bibnamefont{Reitzenstein}},
  \bibinfo{author}{\bibfnamefont{E.}~\bibnamefont{Muljarov}},
  \bibinfo{author}{\bibfnamefont{C.}~\bibnamefont{Kistner}},
  \bibinfo{author}{\bibfnamefont{C.}~\bibnamefont{Schneider}},
  \bibinfo{author}{\bibfnamefont{M.}~\bibnamefont{Strauss}},
  \bibinfo{author}{\bibfnamefont{S.}~\bibnamefont{H{\"o}fling}},
  \bibinfo{author}{\bibfnamefont{A.}~\bibnamefont{Forchel}}, \bibnamefont{and}
  \bibinfo{author}{\bibfnamefont{W.}~\bibnamefont{Langbein}},
  \bibinfo{journal}{Nat. Mater.} \textbf{\bibinfo{volume}{9}},
  \bibinfo{pages}{304} (\bibinfo{year}{2010}).

\bibitem[{\citenamefont{Ota et~al.}(2015)\citenamefont{Ota, Ohta, Kumagai,
  Iwamoto, and Arakawa}}]{ota2015vacuum}
\bibinfo{author}{\bibfnamefont{Y.}~\bibnamefont{Ota}},
  \bibinfo{author}{\bibfnamefont{R.}~\bibnamefont{Ohta}},
  \bibinfo{author}{\bibfnamefont{N.}~\bibnamefont{Kumagai}},
  \bibinfo{author}{\bibfnamefont{S.}~\bibnamefont{Iwamoto}}, \bibnamefont{and}
  \bibinfo{author}{\bibfnamefont{Y.}~\bibnamefont{Arakawa}},
  \bibinfo{journal}{Physical review letters} \textbf{\bibinfo{volume}{114}},
  \bibinfo{pages}{143603} (\bibinfo{year}{2015}).

\bibitem[{\citenamefont{Santori et~al.}(2002)\citenamefont{Santori, Fattal,
  Vu{\v{c}}kovi{\'c}, Solomon, and Yamamoto}}]{santori2002indistinguishable}
\bibinfo{author}{\bibfnamefont{C.}~\bibnamefont{Santori}},
  \bibinfo{author}{\bibfnamefont{D.}~\bibnamefont{Fattal}},
  \bibinfo{author}{\bibfnamefont{J.}~\bibnamefont{Vu{\v{c}}kovi{\'c}}},
  \bibinfo{author}{\bibfnamefont{G.~S.} \bibnamefont{Solomon}},
  \bibnamefont{and} \bibinfo{author}{\bibfnamefont{Y.}~\bibnamefont{Yamamoto}},
  \bibinfo{journal}{Nature} \textbf{\bibinfo{volume}{419}},
  \bibinfo{pages}{594} (\bibinfo{year}{2002}).

\bibitem[{\citenamefont{Patel et~al.}(2010)\citenamefont{Patel, Bennett,
  Farrer, Nicoll, Ritchie, and Shields}}]{patel2010two}
\bibinfo{author}{\bibfnamefont{R.~B.} \bibnamefont{Patel}},
  \bibinfo{author}{\bibfnamefont{A.~J.} \bibnamefont{Bennett}},
  \bibinfo{author}{\bibfnamefont{I.}~\bibnamefont{Farrer}},
  \bibinfo{author}{\bibfnamefont{C.~A.} \bibnamefont{Nicoll}},
  \bibinfo{author}{\bibfnamefont{D.~A.} \bibnamefont{Ritchie}},
  \bibnamefont{and} \bibinfo{author}{\bibfnamefont{A.~J.}
  \bibnamefont{Shields}}, \bibinfo{journal}{Nat. Photonics}
  \textbf{\bibinfo{volume}{4}}, \bibinfo{pages}{632} (\bibinfo{year}{2010}).

\bibitem[{\citenamefont{Lettow et~al.}(2010)\citenamefont{Lettow, Rezus, Renn,
  Zumofen, Ikonen, G{\"o}tzinger, and Sandoghdar}}]{lettow2010quantum}
\bibinfo{author}{\bibfnamefont{R.}~\bibnamefont{Lettow}},
  \bibinfo{author}{\bibfnamefont{Y.}~\bibnamefont{Rezus}},
  \bibinfo{author}{\bibfnamefont{A.}~\bibnamefont{Renn}},
  \bibinfo{author}{\bibfnamefont{G.}~\bibnamefont{Zumofen}},
  \bibinfo{author}{\bibfnamefont{E.}~\bibnamefont{Ikonen}},
  \bibinfo{author}{\bibfnamefont{S.}~\bibnamefont{G{\"o}tzinger}},
  \bibnamefont{and}
  \bibinfo{author}{\bibfnamefont{V.}~\bibnamefont{Sandoghdar}},
  \bibinfo{journal}{Phys. Rev. Lett.} \textbf{\bibinfo{volume}{104}},
  \bibinfo{pages}{123605} (\bibinfo{year}{2010}).

\bibitem[{\citenamefont{Gazzano et~al.}(2013)\citenamefont{Gazzano,
  de~Vasconcellos, Arnold, Nowak, Galopin, Sagnes, Lanco, Lema{\^\i}tre, and
  Senellart}}]{gazzano2013bright}
\bibinfo{author}{\bibfnamefont{O.}~\bibnamefont{Gazzano}},
  \bibinfo{author}{\bibfnamefont{S.~M.} \bibnamefont{de~Vasconcellos}},
  \bibinfo{author}{\bibfnamefont{C.}~\bibnamefont{Arnold}},
  \bibinfo{author}{\bibfnamefont{A.}~\bibnamefont{Nowak}},
  \bibinfo{author}{\bibfnamefont{E.}~\bibnamefont{Galopin}},
  \bibinfo{author}{\bibfnamefont{I.}~\bibnamefont{Sagnes}},
  \bibinfo{author}{\bibfnamefont{L.}~\bibnamefont{Lanco}},
  \bibinfo{author}{\bibfnamefont{A.}~\bibnamefont{Lema{\^\i}tre}},
  \bibnamefont{and}
  \bibinfo{author}{\bibfnamefont{P.}~\bibnamefont{Senellart}},
  \bibinfo{journal}{Nat. Commun.} \textbf{\bibinfo{volume}{4}},
  \bibinfo{pages}{1425} (\bibinfo{year}{2013}).

\bibitem[{\citenamefont{He et~al.}(2013)\citenamefont{He, He, Wei, Wu,
  Atat{\"u}re, Schneider, H{\"o}fling, Kamp, Lu, and Pan}}]{he2013demand}
\bibinfo{author}{\bibfnamefont{Y.-M.} \bibnamefont{He}},
  \bibinfo{author}{\bibfnamefont{Y.}~\bibnamefont{He}},
  \bibinfo{author}{\bibfnamefont{Y.-J.} \bibnamefont{Wei}},
  \bibinfo{author}{\bibfnamefont{D.}~\bibnamefont{Wu}},
  \bibinfo{author}{\bibfnamefont{M.}~\bibnamefont{Atat{\"u}re}},
  \bibinfo{author}{\bibfnamefont{C.}~\bibnamefont{Schneider}},
  \bibinfo{author}{\bibfnamefont{S.}~\bibnamefont{H{\"o}fling}},
  \bibinfo{author}{\bibfnamefont{M.}~\bibnamefont{Kamp}},
  \bibinfo{author}{\bibfnamefont{C.-Y.} \bibnamefont{Lu}}, \bibnamefont{and}
  \bibinfo{author}{\bibfnamefont{J.-W.} \bibnamefont{Pan}},
  \bibinfo{journal}{Nat. Nanotechnol.} \textbf{\bibinfo{volume}{8}},
  \bibinfo{pages}{213} (\bibinfo{year}{2013}).

\bibitem[{\citenamefont{M{\"u}ller et~al.}(2014)\citenamefont{M{\"u}ller,
  Bounouar, J{\"o}ns, Gl{\"a}ssl, and Michler}}]{muller2014demand}
\bibinfo{author}{\bibfnamefont{M.}~\bibnamefont{M{\"u}ller}},
  \bibinfo{author}{\bibfnamefont{S.}~\bibnamefont{Bounouar}},
  \bibinfo{author}{\bibfnamefont{K.~D.} \bibnamefont{J{\"o}ns}},
  \bibinfo{author}{\bibfnamefont{M.}~\bibnamefont{Gl{\"a}ssl}},
  \bibnamefont{and} \bibinfo{author}{\bibfnamefont{P.}~\bibnamefont{Michler}},
  \bibinfo{journal}{Nat. Photonics} \textbf{\bibinfo{volume}{8}},
  \bibinfo{pages}{224} (\bibinfo{year}{2014}).

\bibitem[{\citenamefont{Monniello et~al.}(2014)\citenamefont{Monniello, Reigue,
  Hostein, Lemaitre, Martinez, Grousson, and
  Voliotis}}]{monniello2014indistinguishable}
\bibinfo{author}{\bibfnamefont{L.}~\bibnamefont{Monniello}},
  \bibinfo{author}{\bibfnamefont{A.}~\bibnamefont{Reigue}},
  \bibinfo{author}{\bibfnamefont{R.}~\bibnamefont{Hostein}},
  \bibinfo{author}{\bibfnamefont{A.}~\bibnamefont{Lemaitre}},
  \bibinfo{author}{\bibfnamefont{A.}~\bibnamefont{Martinez}},
  \bibinfo{author}{\bibfnamefont{R.}~\bibnamefont{Grousson}}, \bibnamefont{and}
  \bibinfo{author}{\bibfnamefont{V.}~\bibnamefont{Voliotis}},
  \bibinfo{journal}{Phys. Rev. B} \textbf{\bibinfo{volume}{90}},
  \bibinfo{pages}{041303} (\bibinfo{year}{2014}).

\bibitem[{\citenamefont{Sipahigil et~al.}(2014)\citenamefont{Sipahigil, Jahnke,
  Rogers, Teraji, Isoya, Zibrov, Jelezko, and
  Lukin}}]{sipahigil2014indistinguishable}
\bibinfo{author}{\bibfnamefont{A.}~\bibnamefont{Sipahigil}},
  \bibinfo{author}{\bibfnamefont{K.~D.} \bibnamefont{Jahnke}},
  \bibinfo{author}{\bibfnamefont{L.~J.} \bibnamefont{Rogers}},
  \bibinfo{author}{\bibfnamefont{T.}~\bibnamefont{Teraji}},
  \bibinfo{author}{\bibfnamefont{J.}~\bibnamefont{Isoya}},
  \bibinfo{author}{\bibfnamefont{A.~S.} \bibnamefont{Zibrov}},
  \bibinfo{author}{\bibfnamefont{F.}~\bibnamefont{Jelezko}}, \bibnamefont{and}
  \bibinfo{author}{\bibfnamefont{M.~D.} \bibnamefont{Lukin}},
  \bibinfo{journal}{Phys. Rev. Lett.} \textbf{\bibinfo{volume}{113}},
  \bibinfo{pages}{113602} (\bibinfo{year}{2014}).

\bibitem[{\citenamefont{Wei et~al.}(2014)\citenamefont{Wei, He, Chen, Hu, He,
  Wu, Kamp, Hoefling, Lu, Pan et~al.}}]{wei2014deterministic}
\bibinfo{author}{\bibfnamefont{Y.-J.} \bibnamefont{Wei}},
  \bibinfo{author}{\bibfnamefont{Y.-M.} \bibnamefont{He}},
  \bibinfo{author}{\bibfnamefont{M.-C.} \bibnamefont{Chen}},
  \bibinfo{author}{\bibfnamefont{Y.-N.} \bibnamefont{Hu}},
  \bibinfo{author}{\bibfnamefont{Y.}~\bibnamefont{He}},
  \bibinfo{author}{\bibfnamefont{D.}~\bibnamefont{Wu}},
  \bibinfo{author}{\bibfnamefont{M.}~\bibnamefont{Kamp}},
  \bibinfo{author}{\bibfnamefont{S.}~\bibnamefont{Hoefling}},
  \bibinfo{author}{\bibfnamefont{C.-Y.} \bibnamefont{Lu}},
  \bibinfo{author}{\bibfnamefont{J.-W.} \bibnamefont{Pan}},
  \bibnamefont{et~al.}, \bibinfo{journal}{Nano Lett.}  (\bibinfo{year}{2014}).

\bibitem[{\citenamefont{Giesz et~al.}(2015)\citenamefont{Giesz, Portalupi,
  Grange, Ant\'on, De~Santis, Demory, Somaschi, Sagnes, Lema\^{\i}tre, Lanco
  et~al.}}]{PhysRevB.92.161302}
\bibinfo{author}{\bibfnamefont{V.}~\bibnamefont{Giesz}},
  \bibinfo{author}{\bibfnamefont{S.~L.} \bibnamefont{Portalupi}},
  \bibinfo{author}{\bibfnamefont{T.}~\bibnamefont{Grange}},
  \bibinfo{author}{\bibfnamefont{C.}~\bibnamefont{Ant\'on}},
  \bibinfo{author}{\bibfnamefont{L.}~\bibnamefont{De~Santis}},
  \bibinfo{author}{\bibfnamefont{J.}~\bibnamefont{Demory}},
  \bibinfo{author}{\bibfnamefont{N.}~\bibnamefont{Somaschi}},
  \bibinfo{author}{\bibfnamefont{I.}~\bibnamefont{Sagnes}},
  \bibinfo{author}{\bibfnamefont{A.}~\bibnamefont{Lema\^{\i}tre}},
  \bibinfo{author}{\bibfnamefont{L.}~\bibnamefont{Lanco}},
  \bibnamefont{et~al.}, \bibinfo{journal}{Phys. Rev. B}
  \textbf{\bibinfo{volume}{92}}, \bibinfo{pages}{161302}
  (\bibinfo{year}{2015}).

\bibitem[{\citenamefont{Somaschi et~al.}(2015)\citenamefont{Somaschi, Giesz,
  De~Santis, Loredo, Almeida, Hornecker, Portalupi, Grange, Anton, Demory
  et~al.}}]{somaschi2015near}
\bibinfo{author}{\bibfnamefont{N.}~\bibnamefont{Somaschi}},
  \bibinfo{author}{\bibfnamefont{V.}~\bibnamefont{Giesz}},
  \bibinfo{author}{\bibfnamefont{L.}~\bibnamefont{De~Santis}},
  \bibinfo{author}{\bibfnamefont{J.}~\bibnamefont{Loredo}},
  \bibinfo{author}{\bibfnamefont{M.}~\bibnamefont{Almeida}},
  \bibinfo{author}{\bibfnamefont{G.}~\bibnamefont{Hornecker}},
  \bibinfo{author}{\bibfnamefont{S.}~\bibnamefont{Portalupi}},
  \bibinfo{author}{\bibfnamefont{T.}~\bibnamefont{Grange}},
  \bibinfo{author}{\bibfnamefont{C.}~\bibnamefont{Anton}},
  \bibinfo{author}{\bibfnamefont{J.}~\bibnamefont{Demory}},
  \bibnamefont{et~al.}, \bibinfo{journal}{arXiv preprint arXiv:1510.06499}
  (\bibinfo{year}{2015}).

\bibitem[{\citenamefont{Loredo et~al.}(2016)\citenamefont{Loredo, Zakaria,
  Somaschi, Anton, de~Santis, Giesz, Grange, Broome, Gazzano, Coppola
  et~al.}}]{loredo2016scalable}
\bibinfo{author}{\bibfnamefont{J.~C.} \bibnamefont{Loredo}},
  \bibinfo{author}{\bibfnamefont{N.~A.} \bibnamefont{Zakaria}},
  \bibinfo{author}{\bibfnamefont{N.}~\bibnamefont{Somaschi}},
  \bibinfo{author}{\bibfnamefont{C.}~\bibnamefont{Anton}},
  \bibinfo{author}{\bibfnamefont{L.}~\bibnamefont{de~Santis}},
  \bibinfo{author}{\bibfnamefont{V.}~\bibnamefont{Giesz}},
  \bibinfo{author}{\bibfnamefont{T.}~\bibnamefont{Grange}},
  \bibinfo{author}{\bibfnamefont{M.~A.} \bibnamefont{Broome}},
  \bibinfo{author}{\bibfnamefont{O.}~\bibnamefont{Gazzano}},
  \bibinfo{author}{\bibfnamefont{G.}~\bibnamefont{Coppola}},
  \bibnamefont{et~al.}, \bibinfo{journal}{Optica} \textbf{\bibinfo{volume}{3}},
  \bibinfo{pages}{433} (\bibinfo{year}{2016}).

\bibitem[{\citenamefont{Wilson-Rae and
  Imamo{\u{g}}lu}(2002)}]{wilson2002quantum}
\bibinfo{author}{\bibfnamefont{I.}~\bibnamefont{Wilson-Rae}} \bibnamefont{and}
  \bibinfo{author}{\bibfnamefont{A.}~\bibnamefont{Imamo{\u{g}}lu}},
  \bibinfo{journal}{Physical Review B} \textbf{\bibinfo{volume}{65}},
  \bibinfo{pages}{235311} (\bibinfo{year}{2002}).

\bibitem[{\citenamefont{Kaer et~al.}(2010)\citenamefont{Kaer, Nielsen, Lodahl,
  Jauho, and M{\o}rk}}]{kaer2010non}
\bibinfo{author}{\bibfnamefont{P.}~\bibnamefont{Kaer}},
  \bibinfo{author}{\bibfnamefont{T.~R.} \bibnamefont{Nielsen}},
  \bibinfo{author}{\bibfnamefont{P.}~\bibnamefont{Lodahl}},
  \bibinfo{author}{\bibfnamefont{A.-P.} \bibnamefont{Jauho}}, \bibnamefont{and}
  \bibinfo{author}{\bibfnamefont{J.}~\bibnamefont{M{\o}rk}},
  \bibinfo{journal}{Phys. Rev. Lett.} \textbf{\bibinfo{volume}{104}},
  \bibinfo{pages}{157401} (\bibinfo{year}{2010}).

\bibitem[{\citenamefont{Hohenester}(2010)}]{hohenester2010cavity}
\bibinfo{author}{\bibfnamefont{U.}~\bibnamefont{Hohenester}},
  \bibinfo{journal}{Physical Review B} \textbf{\bibinfo{volume}{81}},
  \bibinfo{pages}{155303} (\bibinfo{year}{2010}).

\bibitem[{\citenamefont{Roy and Hughes}(2011)}]{roy2011phonon}
\bibinfo{author}{\bibfnamefont{C.}~\bibnamefont{Roy}} \bibnamefont{and}
  \bibinfo{author}{\bibfnamefont{S.}~\bibnamefont{Hughes}},
  \bibinfo{journal}{Physical review letters} \textbf{\bibinfo{volume}{106}},
  \bibinfo{pages}{247403} (\bibinfo{year}{2011}).

\bibitem[{\citenamefont{Hughes et~al.}(2011)\citenamefont{Hughes, Yao, Milde,
  Knorr, Dalacu, Mnaymneh, Sazonova, Poole, Aers, Lapointe
  et~al.}}]{hughes2011influence}
\bibinfo{author}{\bibfnamefont{S.}~\bibnamefont{Hughes}},
  \bibinfo{author}{\bibfnamefont{P.}~\bibnamefont{Yao}},
  \bibinfo{author}{\bibfnamefont{F.}~\bibnamefont{Milde}},
  \bibinfo{author}{\bibfnamefont{A.}~\bibnamefont{Knorr}},
  \bibinfo{author}{\bibfnamefont{D.}~\bibnamefont{Dalacu}},
  \bibinfo{author}{\bibfnamefont{K.}~\bibnamefont{Mnaymneh}},
  \bibinfo{author}{\bibfnamefont{V.}~\bibnamefont{Sazonova}},
  \bibinfo{author}{\bibfnamefont{P.}~\bibnamefont{Poole}},
  \bibinfo{author}{\bibfnamefont{G.}~\bibnamefont{Aers}},
  \bibinfo{author}{\bibfnamefont{J.}~\bibnamefont{Lapointe}},
  \bibnamefont{et~al.}, \bibinfo{journal}{Physical Review B}
  \textbf{\bibinfo{volume}{83}}, \bibinfo{pages}{165313}
  (\bibinfo{year}{2011}).

\bibitem[{\citenamefont{Roy-Choudhury and Hughes}(2015)}]{roy2015spontaneous}
\bibinfo{author}{\bibfnamefont{K.}~\bibnamefont{Roy-Choudhury}}
  \bibnamefont{and} \bibinfo{author}{\bibfnamefont{S.}~\bibnamefont{Hughes}},
  \bibinfo{journal}{Optica} \textbf{\bibinfo{volume}{2}}, \bibinfo{pages}{434}
  (\bibinfo{year}{2015}).

\bibitem[{\citenamefont{Hohenester et~al.}(2009)\citenamefont{Hohenester,
  Laucht, Kaniber, Hauke, Neumann, Mohtashami, Seliger, Bichler, and
  Finley}}]{hohenester2009phonon}
\bibinfo{author}{\bibfnamefont{U.}~\bibnamefont{Hohenester}},
  \bibinfo{author}{\bibfnamefont{A.}~\bibnamefont{Laucht}},
  \bibinfo{author}{\bibfnamefont{M.}~\bibnamefont{Kaniber}},
  \bibinfo{author}{\bibfnamefont{N.}~\bibnamefont{Hauke}},
  \bibinfo{author}{\bibfnamefont{A.}~\bibnamefont{Neumann}},
  \bibinfo{author}{\bibfnamefont{A.}~\bibnamefont{Mohtashami}},
  \bibinfo{author}{\bibfnamefont{M.}~\bibnamefont{Seliger}},
  \bibinfo{author}{\bibfnamefont{M.}~\bibnamefont{Bichler}}, \bibnamefont{and}
  \bibinfo{author}{\bibfnamefont{J.~J.} \bibnamefont{Finley}},
  \bibinfo{journal}{Physical Review B} \textbf{\bibinfo{volume}{80}},
  \bibinfo{pages}{201311} (\bibinfo{year}{2009}).

\bibitem[{\citenamefont{Ota et~al.}(2009)\citenamefont{Ota, Iwamoto, Kumagai,
  and Arakawa}}]{ota2009impact}
\bibinfo{author}{\bibfnamefont{Y.}~\bibnamefont{Ota}},
  \bibinfo{author}{\bibfnamefont{S.}~\bibnamefont{Iwamoto}},
  \bibinfo{author}{\bibfnamefont{N.}~\bibnamefont{Kumagai}}, \bibnamefont{and}
  \bibinfo{author}{\bibfnamefont{Y.}~\bibnamefont{Arakawa}},
  \bibinfo{journal}{arXiv preprint arXiv:0908.0788}  (\bibinfo{year}{2009}).

\bibitem[{\citenamefont{Majumdar et~al.}(2011)\citenamefont{Majumdar, Kim,
  Gong, Bajcsy, and Vu{\v{c}}kovi{\'c}}}]{majumdar2011phonon}
\bibinfo{author}{\bibfnamefont{A.}~\bibnamefont{Majumdar}},
  \bibinfo{author}{\bibfnamefont{E.~D.} \bibnamefont{Kim}},
  \bibinfo{author}{\bibfnamefont{Y.}~\bibnamefont{Gong}},
  \bibinfo{author}{\bibfnamefont{M.}~\bibnamefont{Bajcsy}}, \bibnamefont{and}
  \bibinfo{author}{\bibfnamefont{J.}~\bibnamefont{Vu{\v{c}}kovi{\'c}}},
  \bibinfo{journal}{Physical Review B} \textbf{\bibinfo{volume}{84}},
  \bibinfo{pages}{085309} (\bibinfo{year}{2011}).

\bibitem[{\citenamefont{Calic et~al.}(2011)\citenamefont{Calic, Gallo, Felici,
  Atlasov, Dwir, Rudra, Biasiol, Sorba, Tarel, Savona
  et~al.}}]{calic2011phonon}
\bibinfo{author}{\bibfnamefont{M.}~\bibnamefont{Calic}},
  \bibinfo{author}{\bibfnamefont{P.}~\bibnamefont{Gallo}},
  \bibinfo{author}{\bibfnamefont{M.}~\bibnamefont{Felici}},
  \bibinfo{author}{\bibfnamefont{K.}~\bibnamefont{Atlasov}},
  \bibinfo{author}{\bibfnamefont{B.}~\bibnamefont{Dwir}},
  \bibinfo{author}{\bibfnamefont{A.}~\bibnamefont{Rudra}},
  \bibinfo{author}{\bibfnamefont{G.}~\bibnamefont{Biasiol}},
  \bibinfo{author}{\bibfnamefont{L.}~\bibnamefont{Sorba}},
  \bibinfo{author}{\bibfnamefont{G.}~\bibnamefont{Tarel}},
  \bibinfo{author}{\bibfnamefont{V.}~\bibnamefont{Savona}},
  \bibnamefont{et~al.}, \bibinfo{journal}{Physical review letters}
  \textbf{\bibinfo{volume}{106}}, \bibinfo{pages}{227402}
  (\bibinfo{year}{2011}).

\bibitem[{\citenamefont{Madsen et~al.}(2013)\citenamefont{Madsen, Kaer,
  Kreiner-M{\o}ller, Stobbe, Nysteen, M{\o}rk, and
  Lodahl}}]{madsen2013measuring}
\bibinfo{author}{\bibfnamefont{K.~H.} \bibnamefont{Madsen}},
  \bibinfo{author}{\bibfnamefont{P.}~\bibnamefont{Kaer}},
  \bibinfo{author}{\bibfnamefont{A.}~\bibnamefont{Kreiner-M{\o}ller}},
  \bibinfo{author}{\bibfnamefont{S.}~\bibnamefont{Stobbe}},
  \bibinfo{author}{\bibfnamefont{A.}~\bibnamefont{Nysteen}},
  \bibinfo{author}{\bibfnamefont{J.}~\bibnamefont{M{\o}rk}}, \bibnamefont{and}
  \bibinfo{author}{\bibfnamefont{P.}~\bibnamefont{Lodahl}},
  \bibinfo{journal}{Physical Review B} \textbf{\bibinfo{volume}{88}},
  \bibinfo{pages}{045316} (\bibinfo{year}{2013}).

\bibitem[{\citenamefont{Valente et~al.}(2014)\citenamefont{Valente,
  Suffczy{\'n}ski, Jakubczyk, Dousse, Lema{\^\i}tre, Sagnes, Lanco, Voisin,
  Auff{\`e}ves, and Senellart}}]{valente2014frequency}
\bibinfo{author}{\bibfnamefont{D.}~\bibnamefont{Valente}},
  \bibinfo{author}{\bibfnamefont{J.}~\bibnamefont{Suffczy{\'n}ski}},
  \bibinfo{author}{\bibfnamefont{T.}~\bibnamefont{Jakubczyk}},
  \bibinfo{author}{\bibfnamefont{A.}~\bibnamefont{Dousse}},
  \bibinfo{author}{\bibfnamefont{A.}~\bibnamefont{Lema{\^\i}tre}},
  \bibinfo{author}{\bibfnamefont{I.}~\bibnamefont{Sagnes}},
  \bibinfo{author}{\bibfnamefont{L.}~\bibnamefont{Lanco}},
  \bibinfo{author}{\bibfnamefont{P.}~\bibnamefont{Voisin}},
  \bibinfo{author}{\bibfnamefont{A.}~\bibnamefont{Auff{\`e}ves}},
  \bibnamefont{and}
  \bibinfo{author}{\bibfnamefont{P.}~\bibnamefont{Senellart}},
  \bibinfo{journal}{Physical Review B} \textbf{\bibinfo{volume}{89}},
  \bibinfo{pages}{041302} (\bibinfo{year}{2014}).

\bibitem[{\citenamefont{Portalupi et~al.}(2015)\citenamefont{Portalupi,
  Hornecker, Giesz, Grange, Lema{\^\i}tre, Demory, Sagnes, Lanzillotti-Kimura,
  Lanco, Auff{\`e}ves et~al.}}]{Portalupi2015Bright}
\bibinfo{author}{\bibfnamefont{S.~L.} \bibnamefont{Portalupi}},
  \bibinfo{author}{\bibfnamefont{G.}~\bibnamefont{Hornecker}},
  \bibinfo{author}{\bibfnamefont{V.}~\bibnamefont{Giesz}},
  \bibinfo{author}{\bibfnamefont{T.}~\bibnamefont{Grange}},
  \bibinfo{author}{\bibfnamefont{A.}~\bibnamefont{Lema{\^\i}tre}},
  \bibinfo{author}{\bibfnamefont{J.}~\bibnamefont{Demory}},
  \bibinfo{author}{\bibfnamefont{I.}~\bibnamefont{Sagnes}},
  \bibinfo{author}{\bibfnamefont{N.~D.} \bibnamefont{Lanzillotti-Kimura}},
  \bibinfo{author}{\bibfnamefont{L.}~\bibnamefont{Lanco}},
  \bibinfo{author}{\bibfnamefont{A.}~\bibnamefont{Auff{\`e}ves}},
  \bibnamefont{et~al.}, \bibinfo{journal}{Nano letters}
  \textbf{\bibinfo{volume}{15}}, \bibinfo{pages}{6290} (\bibinfo{year}{2015}).

\bibitem[{\citenamefont{Hopfmann et~al.}(2015)\citenamefont{Hopfmann,
  Musia{\l}, Strau{\ss}, Barth, Gl{\"a}ssl, Vagov, Schneider, H{\"o}fling,
  Kamp, Axt et~al.}}]{hopfmann2015compensation}
\bibinfo{author}{\bibfnamefont{C.}~\bibnamefont{Hopfmann}},
  \bibinfo{author}{\bibfnamefont{A.}~\bibnamefont{Musia{\l}}},
  \bibinfo{author}{\bibfnamefont{M.}~\bibnamefont{Strau{\ss}}},
  \bibinfo{author}{\bibfnamefont{A.}~\bibnamefont{Barth}},
  \bibinfo{author}{\bibfnamefont{M.}~\bibnamefont{Gl{\"a}ssl}},
  \bibinfo{author}{\bibfnamefont{A.}~\bibnamefont{Vagov}},
  \bibinfo{author}{\bibfnamefont{C.}~\bibnamefont{Schneider}},
  \bibinfo{author}{\bibfnamefont{S.}~\bibnamefont{H{\"o}fling}},
  \bibinfo{author}{\bibfnamefont{M.}~\bibnamefont{Kamp}},
  \bibinfo{author}{\bibfnamefont{V.}~\bibnamefont{Axt}}, \bibnamefont{et~al.},
  \bibinfo{journal}{Physical Review B} \textbf{\bibinfo{volume}{92}},
  \bibinfo{pages}{245403} (\bibinfo{year}{2015}).

\bibitem[{\citenamefont{M{\"u}ller et~al.}(2015)\citenamefont{M{\"u}ller,
  Fischer, Rundquist, Dory, Lagoudakis, Sarmiento, Kelaita, Borish, and
  Vu{\v{c}}kovi{\'c}}}]{muller2015ultrafast}
\bibinfo{author}{\bibfnamefont{K.}~\bibnamefont{M{\"u}ller}},
  \bibinfo{author}{\bibfnamefont{K.~A.} \bibnamefont{Fischer}},
  \bibinfo{author}{\bibfnamefont{A.}~\bibnamefont{Rundquist}},
  \bibinfo{author}{\bibfnamefont{C.}~\bibnamefont{Dory}},
  \bibinfo{author}{\bibfnamefont{K.~G.} \bibnamefont{Lagoudakis}},
  \bibinfo{author}{\bibfnamefont{T.}~\bibnamefont{Sarmiento}},
  \bibinfo{author}{\bibfnamefont{Y.~A.} \bibnamefont{Kelaita}},
  \bibinfo{author}{\bibfnamefont{V.}~\bibnamefont{Borish}}, \bibnamefont{and}
  \bibinfo{author}{\bibfnamefont{J.}~\bibnamefont{Vu{\v{c}}kovi{\'c}}},
  \bibinfo{journal}{Physical Review X} \textbf{\bibinfo{volume}{5}},
  \bibinfo{pages}{031006} (\bibinfo{year}{2015}).

\bibitem[{\citenamefont{Kaer et~al.}(2012)\citenamefont{Kaer, Nielsen, Lodahl,
  Jauho, and M{\o}rk}}]{kaer2012microscopic}
\bibinfo{author}{\bibfnamefont{P.}~\bibnamefont{Kaer}},
  \bibinfo{author}{\bibfnamefont{T.~R.} \bibnamefont{Nielsen}},
  \bibinfo{author}{\bibfnamefont{P.}~\bibnamefont{Lodahl}},
  \bibinfo{author}{\bibfnamefont{A.-P.} \bibnamefont{Jauho}}, \bibnamefont{and}
  \bibinfo{author}{\bibfnamefont{J.}~\bibnamefont{M{\o}rk}},
  \bibinfo{journal}{Physical Review B} \textbf{\bibinfo{volume}{86}},
  \bibinfo{pages}{085302} (\bibinfo{year}{2012}).

\bibitem[{\citenamefont{Kaer and M{\o}rk}(2014)}]{kaer2014decoherence}
\bibinfo{author}{\bibfnamefont{P.}~\bibnamefont{Kaer}} \bibnamefont{and}
  \bibinfo{author}{\bibfnamefont{J.}~\bibnamefont{M{\o}rk}},
  \bibinfo{journal}{Phys. Rev. B} \textbf{\bibinfo{volume}{90}},
  \bibinfo{pages}{035312} (\bibinfo{year}{2014}).

\bibitem[{\citenamefont{Kaer et~al.}(2013{\natexlab{a}})\citenamefont{Kaer,
  Lodahl, Jauho, and Mork}}]{kaer2013microscopic}
\bibinfo{author}{\bibfnamefont{P.}~\bibnamefont{Kaer}},
  \bibinfo{author}{\bibfnamefont{P.}~\bibnamefont{Lodahl}},
  \bibinfo{author}{\bibfnamefont{A.-P.} \bibnamefont{Jauho}}, \bibnamefont{and}
  \bibinfo{author}{\bibfnamefont{J.}~\bibnamefont{Mork}},
  \bibinfo{journal}{Phys. Rev. B} \textbf{\bibinfo{volume}{87}},
  \bibinfo{pages}{081308} (\bibinfo{year}{2013}{\natexlab{a}}).

\bibitem[{\citenamefont{Mahan}(2000)}]{mahan2000many}
\bibinfo{author}{\bibfnamefont{G.~D.} \bibnamefont{Mahan}},
  \emph{\bibinfo{title}{Many-particle physics}} (\bibinfo{publisher}{Plenum,
  New York}, \bibinfo{year}{2000}).

\bibitem[{\citenamefont{Haug and Jauho}(2008)}]{haug2008quantum}
\bibinfo{author}{\bibfnamefont{H.}~\bibnamefont{Haug}} \bibnamefont{and}
  \bibinfo{author}{\bibfnamefont{A.-P.} \bibnamefont{Jauho}},
  \emph{\bibinfo{title}{Quantum kinetics in transport and optics of
  semiconductors}}, vol. \bibinfo{volume}{123} (\bibinfo{publisher}{Springer},
  \bibinfo{year}{2008}).

\bibitem[{\citenamefont{Auff{\`e}ves et~al.}(2010)\citenamefont{Auff{\`e}ves,
  Gerace, G{\'e}rard, Santos, Andreani, and Poizat}}]{auffeves2010controlling}
\bibinfo{author}{\bibfnamefont{A.}~\bibnamefont{Auff{\`e}ves}},
  \bibinfo{author}{\bibfnamefont{D.}~\bibnamefont{Gerace}},
  \bibinfo{author}{\bibfnamefont{J.-M.} \bibnamefont{G{\'e}rard}},
  \bibinfo{author}{\bibfnamefont{M.~F.} \bibnamefont{Santos}},
  \bibinfo{author}{\bibfnamefont{L.}~\bibnamefont{Andreani}}, \bibnamefont{and}
  \bibinfo{author}{\bibfnamefont{J.-P.} \bibnamefont{Poizat}},
  \bibinfo{journal}{Phys. Rev. B} \textbf{\bibinfo{volume}{81}},
  \bibinfo{pages}{245419} (\bibinfo{year}{2010}).

\bibitem[{\citenamefont{Grange et~al.}(2015)\citenamefont{Grange, Hornecker,
  Hunger, Poizat, G{\'e}rard, Senellart, and Auff{\`e}ves}}]{grange2015cavity}
\bibinfo{author}{\bibfnamefont{T.}~\bibnamefont{Grange}},
  \bibinfo{author}{\bibfnamefont{G.}~\bibnamefont{Hornecker}},
  \bibinfo{author}{\bibfnamefont{D.}~\bibnamefont{Hunger}},
  \bibinfo{author}{\bibfnamefont{J.-P.} \bibnamefont{Poizat}},
  \bibinfo{author}{\bibfnamefont{J.-M.} \bibnamefont{G{\'e}rard}},
  \bibinfo{author}{\bibfnamefont{P.}~\bibnamefont{Senellart}},
  \bibnamefont{and}
  \bibinfo{author}{\bibfnamefont{A.}~\bibnamefont{Auff{\`e}ves}},
  \bibinfo{journal}{Physical Review Letters} \textbf{\bibinfo{volume}{114}},
  \bibinfo{pages}{193601} (\bibinfo{year}{2015}).

\bibitem[{\citenamefont{Muljarov and Zimmermann}(2004)}]{muljarov04}
\bibinfo{author}{\bibfnamefont{E.~A.} \bibnamefont{Muljarov}} \bibnamefont{and}
  \bibinfo{author}{\bibfnamefont{R.}~\bibnamefont{Zimmermann}},
  \bibinfo{journal}{Phys. Rev. Lett.} \textbf{\bibinfo{volume}{93}},
  \bibinfo{pages}{237401} (\bibinfo{year}{2004}).

\bibitem[{\citenamefont{Grange}(2009)}]{grange2009decoherence}
\bibinfo{author}{\bibfnamefont{T.}~\bibnamefont{Grange}},
  \bibinfo{journal}{Phys. Rev. B} \textbf{\bibinfo{volume}{80}},
  \bibinfo{pages}{245310} (\bibinfo{year}{2009}).

\bibitem[{\citenamefont{Goan et~al.}(2011)\citenamefont{Goan, Chen, and
  Jian}}]{goan10}
\bibinfo{author}{\bibfnamefont{H.-S.} \bibnamefont{Goan}},
  \bibinfo{author}{\bibfnamefont{P.-W.} \bibnamefont{Chen}}, \bibnamefont{and}
  \bibinfo{author}{\bibfnamefont{C.-C.} \bibnamefont{Jian}},
  \bibinfo{journal}{The Journal of Chemical Physics}
  \textbf{\bibinfo{volume}{134}}, \bibinfo{eid}{124112} (\bibinfo{year}{2011}).

\bibitem[{\citenamefont{Knill et~al.}(2001)\citenamefont{Knill, Laflamme, and
  Milburn}}]{knill2001scheme}
\bibinfo{author}{\bibfnamefont{E.}~\bibnamefont{Knill}},
  \bibinfo{author}{\bibfnamefont{R.}~\bibnamefont{Laflamme}}, \bibnamefont{and}
  \bibinfo{author}{\bibfnamefont{G.~J.} \bibnamefont{Milburn}},
  \bibinfo{journal}{Nature} \textbf{\bibinfo{volume}{409}}, \bibinfo{pages}{46}
  (\bibinfo{year}{2001}).

\bibitem[{\citenamefont{Kaer et~al.}(2013{\natexlab{b}})\citenamefont{Kaer,
  Gregersen, and Mork}}]{kaer2013role}
\bibinfo{author}{\bibfnamefont{P.}~\bibnamefont{Kaer}},
  \bibinfo{author}{\bibfnamefont{N.}~\bibnamefont{Gregersen}},
  \bibnamefont{and} \bibinfo{author}{\bibfnamefont{J.}~\bibnamefont{Mork}},
  \bibinfo{journal}{New Journal of Physics} \textbf{\bibinfo{volume}{15}},
  \bibinfo{pages}{035027} (\bibinfo{year}{2013}{\natexlab{b}}).

\bibitem[{\citenamefont{Machnikowski and
  Jacak}(2004)}]{machnikowski2004resonant}
\bibinfo{author}{\bibfnamefont{P.}~\bibnamefont{Machnikowski}}
  \bibnamefont{and} \bibinfo{author}{\bibfnamefont{L.}~\bibnamefont{Jacak}},
  \bibinfo{journal}{Physical Review B} \textbf{\bibinfo{volume}{69}},
  \bibinfo{pages}{193302} (\bibinfo{year}{2004}).

\bibitem[{\citenamefont{Vagov et~al.}(2007)\citenamefont{Vagov, Croitoru, Axt,
  Kuhn, and Peeters}}]{vagov2007nonmonotonic}
\bibinfo{author}{\bibfnamefont{A.}~\bibnamefont{Vagov}},
  \bibinfo{author}{\bibfnamefont{M.}~\bibnamefont{Croitoru}},
  \bibinfo{author}{\bibfnamefont{V.~M.} \bibnamefont{Axt}},
  \bibinfo{author}{\bibfnamefont{T.}~\bibnamefont{Kuhn}}, \bibnamefont{and}
  \bibinfo{author}{\bibfnamefont{F.}~\bibnamefont{Peeters}},
  \bibinfo{journal}{Physical review letters} \textbf{\bibinfo{volume}{98}},
  \bibinfo{pages}{227403} (\bibinfo{year}{2007}).

\bibitem[{\citenamefont{Ramsay et~al.}(2010)\citenamefont{Ramsay, Gopal,
  Gauger, Nazir, Lovett, Fox, and Skolnick}}]{ramsay2010damping}
\bibinfo{author}{\bibfnamefont{A.}~\bibnamefont{Ramsay}},
  \bibinfo{author}{\bibfnamefont{A.~V.} \bibnamefont{Gopal}},
  \bibinfo{author}{\bibfnamefont{E.}~\bibnamefont{Gauger}},
  \bibinfo{author}{\bibfnamefont{A.}~\bibnamefont{Nazir}},
  \bibinfo{author}{\bibfnamefont{B.~W.} \bibnamefont{Lovett}},
  \bibinfo{author}{\bibfnamefont{A.}~\bibnamefont{Fox}}, \bibnamefont{and}
  \bibinfo{author}{\bibfnamefont{M.}~\bibnamefont{Skolnick}},
  \bibinfo{journal}{Physical review letters} \textbf{\bibinfo{volume}{104}},
  \bibinfo{pages}{017402} (\bibinfo{year}{2010}).

\bibitem[{\citenamefont{McCutcheon and Nazir}(2010)}]{mccutcheon2010quantum}
\bibinfo{author}{\bibfnamefont{D.~P.} \bibnamefont{McCutcheon}}
  \bibnamefont{and} \bibinfo{author}{\bibfnamefont{A.}~\bibnamefont{Nazir}},
  \bibinfo{journal}{New Journal of Physics} \textbf{\bibinfo{volume}{12}},
  \bibinfo{pages}{113042} (\bibinfo{year}{2010}).

\bibitem[{\citenamefont{Gl{\"a}ssl et~al.}(2011)\citenamefont{Gl{\"a}ssl,
  Croitoru, Vagov, Axt, and Kuhn}}]{glassl2011influence}
\bibinfo{author}{\bibfnamefont{M.}~\bibnamefont{Gl{\"a}ssl}},
  \bibinfo{author}{\bibfnamefont{M.}~\bibnamefont{Croitoru}},
  \bibinfo{author}{\bibfnamefont{A.}~\bibnamefont{Vagov}},
  \bibinfo{author}{\bibfnamefont{V.}~\bibnamefont{Axt}}, \bibnamefont{and}
  \bibinfo{author}{\bibfnamefont{T.}~\bibnamefont{Kuhn}},
  \bibinfo{journal}{Physical Review B} \textbf{\bibinfo{volume}{84}},
  \bibinfo{pages}{125304} (\bibinfo{year}{2011}).

\bibitem[{\citenamefont{Debnath et~al.}(2012)\citenamefont{Debnath, Meier,
  Chatel, and Amand}}]{debnath2012chirped}
\bibinfo{author}{\bibfnamefont{A.}~\bibnamefont{Debnath}},
  \bibinfo{author}{\bibfnamefont{C.}~\bibnamefont{Meier}},
  \bibinfo{author}{\bibfnamefont{B.}~\bibnamefont{Chatel}}, \bibnamefont{and}
  \bibinfo{author}{\bibfnamefont{T.}~\bibnamefont{Amand}},
  \bibinfo{journal}{Physical Review B} \textbf{\bibinfo{volume}{86}},
  \bibinfo{pages}{161304} (\bibinfo{year}{2012}).

\bibitem[{\citenamefont{Manson et~al.}(2016)\citenamefont{Manson,
  Roy-Choudhury, and Hughes}}]{manson2016polaron}
\bibinfo{author}{\bibfnamefont{R.}~\bibnamefont{Manson}},
  \bibinfo{author}{\bibfnamefont{K.}~\bibnamefont{Roy-Choudhury}},
  \bibnamefont{and} \bibinfo{author}{\bibfnamefont{S.}~\bibnamefont{Hughes}},
  \bibinfo{journal}{Physical Review B} \textbf{\bibinfo{volume}{93}},
  \bibinfo{pages}{155423} (\bibinfo{year}{2016}).

\bibitem[{\citenamefont{McCutcheon}(2016)}]{mccutcheon2016optical}
\bibinfo{author}{\bibfnamefont{D.~P.} \bibnamefont{McCutcheon}},
  \bibinfo{journal}{Physical Review A} \textbf{\bibinfo{volume}{93}},
  \bibinfo{pages}{022119} (\bibinfo{year}{2016}).

\bibitem[{\citenamefont{Nazir and McCutcheon}(2016)}]{nazir2016modelling}
\bibinfo{author}{\bibfnamefont{A.}~\bibnamefont{Nazir}} \bibnamefont{and}
  \bibinfo{author}{\bibfnamefont{D.~P.} \bibnamefont{McCutcheon}},
  \bibinfo{journal}{Journal of Physics: Condensed Matter}
  \textbf{\bibinfo{volume}{28}}, \bibinfo{pages}{103002}
  (\bibinfo{year}{2016}).

\bibitem[{\citenamefont{Giesz et~al.}(2016)\citenamefont{Giesz, Somaschi,
  Hornecker, Grange, Reznychenko, De~Santis, Demory, Gomez, Sagnes,
  Lema{\^\i}tre et~al.}}]{giesz2016coherent}
\bibinfo{author}{\bibfnamefont{V.}~\bibnamefont{Giesz}},
  \bibinfo{author}{\bibfnamefont{N.}~\bibnamefont{Somaschi}},
  \bibinfo{author}{\bibfnamefont{G.}~\bibnamefont{Hornecker}},
  \bibinfo{author}{\bibfnamefont{T.}~\bibnamefont{Grange}},
  \bibinfo{author}{\bibfnamefont{B.}~\bibnamefont{Reznychenko}},
  \bibinfo{author}{\bibfnamefont{L.}~\bibnamefont{De~Santis}},
  \bibinfo{author}{\bibfnamefont{J.}~\bibnamefont{Demory}},
  \bibinfo{author}{\bibfnamefont{C.}~\bibnamefont{Gomez}},
  \bibinfo{author}{\bibfnamefont{I.}~\bibnamefont{Sagnes}},
  \bibinfo{author}{\bibfnamefont{A.}~\bibnamefont{Lema{\^\i}tre}},
  \bibnamefont{et~al.}, \bibinfo{journal}{Nature Communications}
  \textbf{\bibinfo{volume}{7}} (\bibinfo{year}{2016}).

\bibitem[{\citenamefont{Johnson et~al.}(2015)\citenamefont{Johnson, Dolan,
  Grange, Trichet, Hornecker, Chen, Weng, Hughes, Watt, Auff{\`e}ves
  et~al.}}]{johnson2015tunable}
\bibinfo{author}{\bibfnamefont{S.}~\bibnamefont{Johnson}},
  \bibinfo{author}{\bibfnamefont{P.}~\bibnamefont{Dolan}},
  \bibinfo{author}{\bibfnamefont{T.}~\bibnamefont{Grange}},
  \bibinfo{author}{\bibfnamefont{A.}~\bibnamefont{Trichet}},
  \bibinfo{author}{\bibfnamefont{G.}~\bibnamefont{Hornecker}},
  \bibinfo{author}{\bibfnamefont{Y.}~\bibnamefont{Chen}},
  \bibinfo{author}{\bibfnamefont{L.}~\bibnamefont{Weng}},
  \bibinfo{author}{\bibfnamefont{G.}~\bibnamefont{Hughes}},
  \bibinfo{author}{\bibfnamefont{A.}~\bibnamefont{Watt}},
  \bibinfo{author}{\bibfnamefont{A.}~\bibnamefont{Auff{\`e}ves}},
  \bibnamefont{et~al.}, \bibinfo{journal}{New Journal of Physics}
  \textbf{\bibinfo{volume}{17}}, \bibinfo{pages}{122003}
  (\bibinfo{year}{2015}).

\end{thebibliography}

\end{document}